\pdfoutput=1 
\documentclass[useAMS,usenatbib]{mnras}

\usepackage{graphicx}
\usepackage{times}

\usepackage{eso-pic}

\usepackage{hyperref}
\usepackage{amssymb}
\usepackage{amsmath}
\usepackage{lineno}
\usepackage{setspace}

\newcommand{\mpcoh}{\,h^{-1}\,{\rm Mpc}}

\newcommand{\mocks}{{\rm {DES-BAO-MOCKS}}}

\newcommand{\methodACF}{{\rm {DES-BAO-$\theta$-METHOD}}}
\newcommand{\methodcl}{{\rm {DES-BAO-$\ell$-METHOD}}}
\newcommand{\methodXi}{{\rm {DES-BAO-$s_{\perp}$-METHOD}}}
\newcommand{\photoz}{{\rm {DES-BAO-PHOTOZ}}}
\newcommand{\sample}{{\rm {DES-BAO-SAMPLE}}}

\begin{document}

  \AddToShipoutPictureBG*{%
    \AtPageUpperLeft{%
      \hspace{0.75\paperwidth}%
      \raisebox{-2.5\baselineskip}{%
        \makebox[0pt][l]{\textnormal{DES-2017-0283}}
  }}}%

  \AddToShipoutPictureBG*{%
    \AtPageUpperLeft{%
      \hspace{0.75\paperwidth}%
      \raisebox{-3.5\baselineskip}{%
        \makebox[0pt][l]{\textnormal{FERMILAB-PUB-17-586}}
  }}}%

\title[DES Y1 BAO Measurements] 
{Dark Energy Survey Year 1 Results: Measurement of the Baryon Acoustic Oscillation scale in the distribution of galaxies to redshift 1}


\author[DES Collaboration]{
\parbox{\textwidth}{
\Large
T.~M.~C.~Abbott$^{1}$,
F.~B.~Abdalla$^{2,3}$,
A.~Alarcon$^{4}$,
S.~Allam$^{5}$,
F.~Andrade-Oliveira$^{6,7}$,
J.~Annis$^{5}$,
S.~Avila$^{8,9}$,
M.~Banerji$^{10,11}$,
N.~Banik$^{12,13,14,5}$,
K.~Bechtol$^{15}$
% ...
\begin{center} (The Dark Energy Survey Collaboration) \end{center}
}
\vspace{0.4cm}
\\
}



\date{Prepared for submission to MNRAS}

\pagerange{\pageref{firstpage}--\pageref{lastpage}} \pubyear{2017}
\maketitle
\thanks{For correspondence use des-publication-queries@fnal.gov}
\label{firstpage}

\begin{abstract}
We present angular diameter distance measurements obtained by locating the BAO scale in the distribution of galaxies selected from the first year of Dark Energy Survey data. We consider a sample of over 1.3 million galaxies distributed over a footprint of 1336 deg$^2$ with $0.6 < z_{\rm photo} < 1$ and a typical redshift uncertainty of $0.03(1+z)$. This sample was selected, as fully described in a companion paper,
using a color/magnitude selection that optimizes trade-offs between number density and redshift uncertainty. We investigate the BAO signal in the projected clustering using three conventions, the angular separation, the co-moving transverse separation, and spherical harmonics. Further, we compare results obtained from template based and machine learning photometric redshift determinations. We use 1800 simulations that approximate our sample in order to produce covariance matrices and allow us to validate our distance scale measurement methodology. We measure the angular diameter distance, $D_A$, at the effective redshift of our sample divided by the true physical scale of the BAO feature, $r_{\rm d}$. We obtain close to a 4 per cent distance measurement of $D_A(z_{\rm eff}=0.81)/r_{\rm d} = 10.75\pm 0.43 $. These results are consistent with the flat $\Lambda$CDM concordance cosmological model supported by numerous other recent experimental results. \end{abstract}

\begin{keywords}
  cosmology: observations - (cosmology:) large-scale structure of Universe
\end{keywords}

\clearpage
\begin{tiny}
\begin{spacing}{1.0}
\noindent
$^{1}$ Cerro Tololo Inter-American Observatory, National Optical Astronomy Observatory, Casilla 603, La Serena, Chile\\
$^{2}$ Department of Physics \& Astronomy, University College London, Gower Street, London, WC1E 6BT, UK\\
$^{3}$ Department of Physics and Electronics, Rhodes University, PO Box 94, Grahamstown, 6140, South Africa\\
$^{4}$ Institute of Space Sciences (ICE, CSIC) \& Institut d'Estudis Espacials de Catalunya (IEEC), Campus UAB, Carrer de Can Magrans, s/n,  08193 Barcelona, Spain\\
$^{5}$ Fermi National Accelerator Laboratory, P. O. Box 500, Batavia, IL 60510, USA\\
$^{6}$ Laborat\'orio Interinstitucional de e-Astronomia - LIneA, Rua Gal. Jos\'e Cristino 77, Rio de Janeiro, RJ - 20921-400, Brazil\\
$^{7}$ Instituto de Fisica Teorica, Universidade Estadual Paulista, Sao Paulo, Brazil\\
$^{8}$ Institute of Cosmology \& Gravitation, University of Portsmouth, Portsmouth, PO1 3FX, UK\\
$^{9}$ Instituto de Fisica Teorica UAM/CSIC, Universidad Autonoma de Madrid, 28049 Madrid, Spain\\
$^{10}$ Institute of Astronomy, University of Cambridge, Madingley Road, Cambridge CB3 0HA, UK\\
$^{11}$ Kavli Institute for Cosmology, University of Cambridge, Madingley Road, Cambridge CB3 0HA, UK\\
$^{12}$ Department of Physics, University of Florida, Gainesville, Florida 32611, USA\\
$^{13}$ GRAPPA, Institute of Theoretical Physics, University of Amsterdam, Science Park 904, 1090 GL Amsterdam\\
$^{14}$ Lorentz Institute, Leiden University, Niels Bohrweg 2, Leiden, NL-2333 CA, The Netherlands\\
$^{15}$ LSST, 933 North Cherry Avenue, Tucson, AZ 85721, USA\\
$^{16}$ Observatories of the Carnegie Institution of Washington, 813 Santa Barbara St., Pasadena, CA 91101, USA\\
$^{17}$ Department of Physics and Astronomy, University of Pennsylvania, Philadelphia, PA 19104, USA\\
$^{18}$ CNRS, UMR 7095, Institut d'Astrophysique de Paris, F-75014, Paris, France\\
$^{19}$ Sorbonne Universit\'es, UPMC Univ Paris 06, UMR 7095, Institut d'Astrophysique de Paris, F-75014, Paris, France\\
$^{20}$ SLAC National Accelerator Laboratory, Menlo Park, CA 94025, USA\\
$^{21}$ Kavli Institute for Particle Astrophysics \& Cosmology, P. O. Box 2450, Stanford University, Stanford, CA 94305, USA\\
$^{22}$ Departamento de F\'isica Matem\'atica, Instituto de F\'isica, Universidade de S\~ao Paulo, CP 66318, S\~ao Paulo, SP, 05314-970, Brazil\\
$^{23}$ Observat\'orio Nacional, Rua Gal. Jos\'e Cristino 77, Rio de Janeiro, RJ - 20921-400, Brazil\\
$^{24}$ National Center for Supercomputing Applications, 1205 West Clark St., Urbana, IL 61801, USA\\
$^{25}$ Department of Astronomy, University of Illinois at Urbana-Champaign, 1002 W. Green Street, Urbana, IL 61801, USA\\
$^{26}$ Institut de F\'{\i}sica d'Altes Energies (IFAE), The Barcelona Institute of Science and Technology, Campus UAB, 08193 Bellaterra (Barcelona) Spain\\
$^{27}$ Kavli Institute for Cosmological Physics, University of Chicago, Chicago, IL 60637, USA\\
$^{28}$ School of Physics and Astronomy, Sun Yat-Sen University, Guangzhou 510275, China\\
$^{29}$ Centro de Investigaciones Energ\'eticas, Medioambientales y Tecnol\'ogicas (CIEMAT), Madrid, Spain\\
$^{30}$ George P. and Cynthia Woods Mitchell Institute for Fundamental Physics and Astronomy, and Department of Physics and Astronomy, Texas A\&M University, College Station, TX 77843,  USA\\
$^{31}$ Department of Physics, IIT Hyderabad, Kandi, Telangana 502285, India\\
$^{32}$ Jet Propulsion Laboratory, California Institute of Technology, 4800 Oak Grove Dr., Pasadena, CA 91109, USA\\
$^{33}$ Department of Astronomy/Steward Observatory, 933 North Cherry Avenue, Tucson, AZ 85721-0065, USA\\
$^{34}$ Jodrell Bank Center for Astrophysics, School of Physics and Astronomy, University of Manchester, Oxford Road, Manchester, M13 9PL, UK\\
$^{35}$ Department of Astronomy, University of Michigan, Ann Arbor, MI 48109, USA\\
$^{36}$ Department of Physics, University of Michigan, Ann Arbor, MI 48109, USA\\
$^{37}$ Universit\"ats-Sternwarte, Fakult\"at f\"ur Physik, Ludwig-Maximilians Universit\"at M\"unchen, Scheinerstr. 1, 81679 M\"unchen, Germany\\
$^{38}$ Department of Physics, ETH Zurich, Wolfgang-Pauli-Strasse 16, CH-8093 Zurich, Switzerland\\
$^{39}$ Santa Cruz Institute for Particle Physics, Santa Cruz, CA 95064, USA\\
$^{40}$ Department of Physics, The Ohio State University, Columbus, OH 43210, USA\\
$^{41}$ Center for Cosmology and Astro-Particle Physics, The Ohio State University, Columbus, OH 43210, USA\\
$^{42}$ Max Planck Institute for Extraterrestrial Physics, Giessenbachstrasse, 85748 Garching, Germany\\
$^{43}$ Event Horizon Telescope, Harvard-Smithsonian Center for Astrophysics, MS-42, 60 Garden Street, Cambridge, MA 02138\\
$^{44}$ Australian Astronomical Observatory, North Ryde, NSW 2113, Australia\\
$^{45}$ Argonne National Laboratory, 9700 South Cass Avenue, Lemont, IL 60439, USA\\
$^{46}$ D\'epartement de Physique Th\'eorique and Center for Astroparticle Physics, Universit\'e de Gene\'ve, 24 quai Ernest Ansermet, 1211, Geneva, Switzerland\\
$^{47}$ Department of Astronomy, The Ohio State University, Columbus, OH 43210, USA\\
$^{48}$ Department of Astrophysical Sciences, Princeton University, Peyton Hall, Princeton, NJ 08544, USA\\
$^{49}$ Instituci\'o Catalana de Recerca i Estudis Avan\c{c}ats, E-08010 Barcelona, Spain\\
$^{50}$ Faculty of Physics, Ludwig-Maximilians-Universit\"at, Scheinerstr. 1, 81679 Munich, Germany\\
$^{51}$ Excellence Cluster Universe, Boltzmannstr.\ 2, 85748 Garching, Germany\\
$^{52}$ Department of Physics and Astronomy, Pevensey Building, University of Sussex, Brighton, BN1 9QH, UK\\
$^{53}$ ICTP South American Institute for Fundamental Research\\ Instituto de F\'{\i}sica Te\'orica, Universidade Estadual Paulista, S\~ao Paulo, Brazil\\
$^{54}$ Department of Physics, University of Arizona, Tucson, AZ 85721, USA\\
$^{55}$ Instituto de F\'\i sica, UFRGS, Caixa Postal 15051, Porto Alegre, RS - 91501-970, Brazil\\
$^{56}$ Brookhaven National Laboratory, Bldg 510, Upton, NY 11973, USA\\
$^{57}$ School of Physics and Astronomy, University of Southampton,  Southampton, SO17 1BJ, UK\\
$^{58}$ Instituto de F\'isica Gleb Wataghin, Universidade Estadual de Campinas, 13083-859, Campinas, SP, Brazil\\
$^{59}$ Computer Science and Mathematics Division, Oak Ridge National Laboratory, Oak Ridge, TN 37831\\
$^{60}$ Department of Physics, Stanford University, 382 Via Pueblo Mall, Stanford, CA 94305, USA\\
\end{spacing}
\end{tiny}

\section{Introduction}
The signature of Baryon Acoustic Oscillations (BAO) can be observed in the distribution of tracers of the matter density field and used to measure the expansion history of the Universe. BAO data alone prefer dark energy at greater than 6$\sigma$ and are consistent with a flat $\Lambda$CDM cosmology with $\Omega_{\rm matter} = 0.3$ \citep{Ata17}. A large number of spectroscopic surveys have measured BAO in the distributions of galaxies, quasars, and the the Lyman-$\alpha$ forest, including the Sloan Digital Sky Survey (SDSS) I, II, III, and IV \citealt{Eisenstein05,Gaz09,Percival10,Ross15,Acacia,Ata17,Delubac15,Bautista17}, the 2-degree Field Galaxy Redshift Survey (2dFGRS) \citep{Percival01,2dF}, WiggleZ \citep{Blake11}, and the 6-degree Field Galaxy Survey (6dFGS) \citep{6dF}.

BAO can also be measured using purely photometric data, though at less significance. Radial distance measurements are severely hampered, but some information about the angular diameter distance $D_A$ is still accessible. Analytic analysis of the expected signal is presented in  \cite{SeoEis03,BlakeBridle05,Zhan09}. Recently, \cite{RossOpt3D} (hereafter \methodXi) studied the signal expected to be present for data similar to DES Y1 and recommended the use of the projected two point correlation function, $\xi(s_{\perp})$, as a clustering estimator ideal for extracting the BAO signal. Measurements of the BAO signal in various photometric data samples have been presented in \citealt{Pad07,Estrada09,Hutsi10,Sanchez11,Crocce11,Seo12,Carnero12}, using a variety of methodologies .

We use imaging data from the first year (Y1) of Dark Energy Survey (DES) observations to measure the angular diameter distance to red galaxies with photometric redshifts $0.6 < z_{\rm photo} < 1.0$. DES is a five year program to image a 5,000 deg$^2$ footprint of the Southern hemisphere using five passbands, $grizY$. It will measure the properties of over 300 million galaxies. Here, we use 1.3 million galaxies over 1336 deg$^2$ color and magnitude selected to balance trade-offs in BAO measurement between the redshift precision and the number density. We use these data, supported by 1800 mock realizations of our sample, to allow us to make the first BAO measurement using galaxies centered at $z>0.8$.

The measurement we present is supported by a series of companion papers. \cite{Crocce17} presents the selection of our DES Y1 sample, optimized for $z > 0.6$ BAO measurements, tests of its basic properties, and redshift validation; we denote it \sample  ~hereafter. \cite{Avila17} describes how 1800 realizations approximating the spatial properties of the DES Y1 data sample were produced and validated; we denote it \mocks. Using these mock Y1 realizations, \cite{whetabao} validates and optimizes the methodology for measuring BAO from the angular two point correlation function, $w(\theta)$; we denote it \methodACF.  Analysis of the angular power spectrum is presented in $C_{\ell}$ (\citealt{clbao}; \methodcl). 

In this paper, we collate the results of the above papers. With this basic framework, we identify the BAO signature in the DES Y1 data, and use it to place constraints on the comoving angular diameter distance to the effective redshift of our sample, $z_{\rm eff} = 0.81$. The cosmological implications of this measurement are then discussed. Section \ref{sec:data} summarizes the data we use, including all of its basic properties and details on the mock realizations of the data (mocks). Section \ref{sec:analysis} presents the basic techniques we apply to measure the clustering of galaxies, estimate covariance matrices in order to extract parameter likelihoods, and extract the BAO scale distance from the measurements. Section \ref{sec:mocktest} summarizes tests performed on the 1800 mock Y1 realizations, which help set our fiducial analysis choices. Section \ref{sec:results} presents the clustering measurements and the BAO scale we extract from them. Section \ref{sec:cosmo} compares our measurement to predictions of the flat $\Lambda$CDM model and other BAO scale distance measurements. We conclude in Section 7 with a discussion of future prospects. 

The fiducial cosmology we use for this work is a flat $\Lambda$CDM with $\Omega_{\rm matter} = 0.25$ with $h=0.7$. Such a low matter density is ruled out by current observational constraints (see, e.g., \citealt{Planck2015}). However, the cosmology we use is matched to that of the MICE \citep{CrocceMICE,FosalbaMICE,FosalbaMICE2,CarreteroMICE} $N$-Body simulation, which was used to calibrate the mock galaxy samples we employ to test and validate our methodology. We will demonstrate that our results are not sensitive to this choice.

We note that the determination of the color and magnitude cuts, as well as the overall redshift range and the observational systematics 
treatment, was completed prior to any actual clustering measurement and based on considerations of photo-$z$ performance, area
and number density. Therefore, our sample selection was blind to any potential BAO detection in the data.

\section{Data}
\label{sec:data}
\subsection{DES Year 1 Data}
``DES Year 1'' (Y1) data were obtained in the period of time between August 31, 2013 and February 9, 2014 using the 570-megapixel Dark Energy Camera (DECam; \citealt{DECam}). Y1 contains images occupying a total footprint of more than 1800 deg$^2$ in $grizY$ photometric passbands \citep{DESY1}. The DES Data Management (DESDM) system \citep{DESDM1,DESDM2,DESDM3} detrended, calibrated, and coadded these DES images in order to catalog astrophysical objects. From these results the Y1 `Gold' catalog was produced, which provided photometry and `clean' galaxy samples, as described in \cite{Y1Gold}. The observed footprint is defined by a {\sc Healpix} \citep{healpix} map at resolution $N_{\mathrm{side}}=4096$ and includes only area with a minimum total exposure time of at least 90 seconds in each of the $griz$ bands and a valid calibration solution (see \citealt{Y1Gold} for details). A series of veto masks, covering regions of poor quality or foregrounds for the galaxy sample, reduce the area to 1336~deg$^2$ suitable for LSS study. We describe the additional masks we apply to the data in Section \ref{sec:mask}.

\subsection{BAO Sample Selection}
\label{sec:benchred}

We use a sample selected from the DES Y1 Gold catalog to provide the best BAO constraints. The sample balances number density and photometric redshift uncertainty considerations in order to produce an optimal sample. The sample definition is fully described in \cite{Crocce17} (\sample). We repeat vital information here.

We select our sample based on color, magnitude, and redshift cuts. The primary redshift, colour, and magnitude cuts are
\begin{eqnarray}
17.5 < i_{\rm auto} < 19.0 + 3.0z_{\rm BPZ-AUTO}\label{eq:zcut}\\
(i_{\rm auto} - z_{\rm auto}) + 2.0(r_{\rm auto} - i_{\rm auto}) > 1.7 \label{eq:colorcut}\\
0.6 < z_{\rm photo} < 1.0.
\end{eqnarray}
The color and magnitude cuts use \texttt{mag\_auto} defined in Y1 Gold and $z_{\rm BPZ-AUTO}$ is the BPZ photometric redshift \citep{Benitez00} determined with the \texttt{mag\_auto} photometry.\footnote{While we use \texttt{mag\_auto} photometry for galaxy selection, our redshift estimates rely on a proper multi-object fitting procedure.  The use of \texttt{mag\_auto} for galaxy selection reflects that the latter color measurements only became available after the galaxy selection had already been finalized. } The quantity $z_{\rm photo}$ is the photometric redshift used for a particular sample, we describe these in Section \ref{sec:zphot}.

Stars are removed via the cut 
\begin{equation}
{\rm spread\_model_i} + (5.0/3.0){\rm spreaderr\_model_i} > 0.007
\end{equation}
and we also remove outliers in colour space via
\begin{eqnarray}
-1<g_{\rm auto} - r_{\rm auto}<3\\
-1<r_{\rm auto} - i_{\rm auto}<2.5\\
-1<i_{\rm auto} - z_{\rm auto}<2.
\end{eqnarray}

\begin{table}
\centering
\caption{Characteristics of the DES Y1 BAO sample, as a function of redshift: number of galaxies, redshift uncertainties and fraction of star contamination. Results are shown for the DNF redshift estimate, with BPZ results in parentheses. We used $z$ to denote the mean redshift of the given estimator (as each galaxy has a redshift likelihood). }
\begin{tabular}{lccc}
\hline
\hline
$z_{\rm photo}$ & $N_{\rm gal}$ & $\sigma_{68}/(1+z)$ & $f_{\rm star}$\\
\hline
$0.6 < z < 0.7$ & 386057 (332242) & 0.023 (0.027) & 0.004 (0.018) \\
$0.7 < z < 0.8$ & 353789 (429366) & 0.028 (0.031) & 0.037 (0.042)\\
$0.8 < z < 0.9$ & 330959 (380059) & 0.029 (0.034) & 0.012 (0.015)\\
$0.9 < z < 1.0$ & 229395 (180560) & 0.036 (0.039) & 0.015 (0.006)\\
\hline
\hline
\label{tab:sample}
\end{tabular}
\end{table}

\subsection{Redshifts}
\label{sec:zphot}

We define two samples based on two different photometric redshift algorithms, BPZ and DNF \citep{DeVicente15}. For both samples, our point estimates of the redshift, $z_{\rm photo}$, use photozs determined using the `MOF' photometry defined in Y1 Gold. However, for each sample, we use the BPZ value calculated using \texttt{mag\_auto} photometry with the sample selection cut defined in Eq. \ref{eq:zcut}; this is the only time that photozs estimated using \texttt{mag\_auto} photometry are used. We use the DNF method as our fiducial redshift estimator as it performed better, in terms of both precision and accuracy, on validation tests (see \sample ~ and \photoz ~ for more details). As a robustness check, we also compare our results to those derived using BPZ redshifts, as determined using MOF photometry. In Table \ref{tab:sample} we present the statistics of each sample (after masking, see Section \ref{sec:mask}) divided into redshift bins of width $\Delta z = 0.1$. We define $\sigma_{68}$ as the half width of the interval containing the median $68\%$ of values in the distribution of $(z_{photo}-z_{true})/(1 +z_{true})$. This is estimated based on the redshift validation described in \sample ~and \photoz.

The redshift estimate for each individual galaxy has substantial uncertainty, typically with non-Gaussian likelihood distributions for $z_{\rm true}$. We use individual point estimates of the redshift both for binning in redshift and for calculating transverse separation in $h^{-1}$Mpc. In order to do so, we use the mean redshift produced by the given redshift estimator. In what follows, we will refer to this estimate of the redshift by $z$ (dropping the ${\rm photo}$ sub-script).

Plots for the estimated true redshift distribution in each of the tomographic bins given in Table \ref{tab:sample} are presented in \sample. For indicative purposes we quote here estimates of the fraction of galaxies assigned to a redshift bin via their photo-$z$ that would actually lie in a different bin. For the first tomographic bin $30\%$ for the galaxies are estimated to migrate from the adjacent bins, for the second and third redshift bin this number increases to $40\%$ and for the last bin it is $\%50$.

We determine an effective redshift for our sample of $z_{\rm eff} = 0.81$. This is determined from the mean redshift obtained when applying all weights, including those defined in Eq. \ref{eq:xiFKP}, which account for the expected signal to noise as a function of redshift. See \sample ~and \photoz ~for further details on the redshifts used for the DES Y1 BAO sample and their validation.

\subsection{Mask}
\label{sec:mask}
\begin{figure}
\includegraphics[width=84mm]{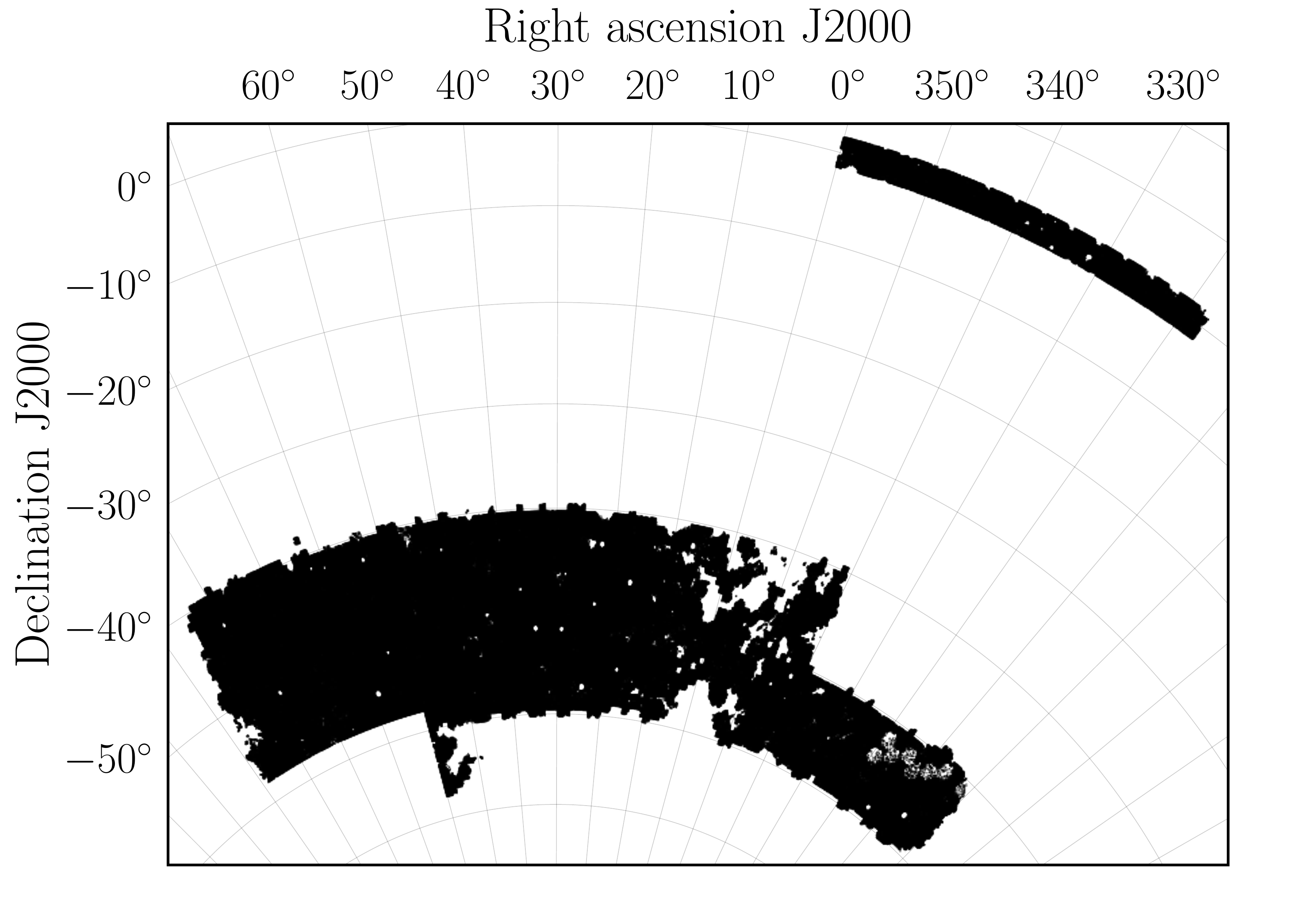}
  \caption{The black shaded region represents the area on the sky to which we restrict our DES Y1 BAO analysis. See Section \ref{sec:mask}.}
  \label{fig:footprint}
\end{figure}

The most basic requirement is that DES Y1 observations exist in $griz$, since our selection requires each of the four bands. We use the Y1 Gold coverage maps, at \textsc{Healpix} resolution $N_{\rm side} =4096$, to enforce this condition. We require that any pixel be at least 80 per cent covered in the four bands simultaneously. The minimum coverage across all four bands is then used as a weight for the pixel. We also require that the depth limit in each band is at the level required of our colour/magnitude selection.
The Y1 Gold catalog includes 10$\sigma$ \texttt{MAG\_AUTO} depth maps for each band, again at \textsc{Healpix} resolution $N_{\rm side} =4096$. This is an angular size of 0.014 degrees and less than one tenth of the resolution of any clustering statistics we employ. We consider only areas with $i$ band depth greater than 22 and depth in the other bands great enough to reliably measure the colour defined by Eq. \ref{eq:colorcut}. This involved removing regions of the footprint that did not fufill the condition $(2 r_{\rm maglim} - z_{\rm maglim}) < 23.7$. We further cut out `bad regions' identified in Y1GOLD (removing everything with flag bit $>$ 2 in their table 5), and areas with $z$-band seeing greater than 3.7 pixels. We also remove a patch with area 18 deg$^2$, where the airmass quantity was corrupted. The resulting footprint occupies 1336 deg$^2$ and is shown in Fig. \ref{fig:footprint}. 

\subsection{Observational Systematics}
\label{sec:obsys}

As detailed in \sample, we have found significant dependencies between the number density of galaxies in our sample and three observational quantities: the local stellar density, the PSF FWHM (`seeing'), and the detection limit (depth). The dependency with stellar density is understood as stellar contamination: some fraction of our ``galaxies'' are in fact stars. The inferred stellar contamination, $f_{\rm star}$, is listed in Table \ref{tab:sample}. The dependencies with seeing and depth are similar to what was found for a separate DES Y1 sample by \cite{ElvinPoole17}. We correct for the systematic dependencies via weights that we assign to the catalog, which when applied remove the trend with the quantity in question. The total weight, $w_{\rm sys}$, is the product of all three individual weights. We apply $w_{\rm sys}$ to all counting and clustering statistics presented in this paper, except where we omit it as a test of robustness. See \sample ~for full details on the construction of the weights. We find that the weights have a minimal impact on our analysis.

\subsection{Mocks}
\label{sec:mocks}
We simulate our sample using 1800 mock DES Y1 catalogs. These are fully described in \cite{Avila17} (\mocks) and we only repeat the basic details here. Each mock matches the footprint, clustering, and redshift accuracy/distribution of our DES sample. Halo catalogs are generated using the \textsc{halogen}  technique \citep{halogen}, based on a 2LPT density field with an exponential bias. The method is tuned to reproduce the halo clustering as a function of mass and redshift of a reference $N$-Body simulation (MICE; \citealt{FosalbaMICE}). We use a box size of $L_{\rm box}=3072h^{-1}$Mpc with a halo mass resolution of $M_h=2.5\times 10^{12}h^{-1}M_\odot $. Haloes are then arranged in a lightcone by superposition of 11 snapshots. We tile 8 replicas of the box together using the periodic conditions to construct a full sky mock from $z=0$ to $z=1.42$, from which we draw eight mock catalogues. 

We take care to properly reproduce the redshift properties of our DES Y1 sample. For each mock galaxy, we have the true redshift $z_{\rm true}$. We thus require an estimate of the joint distribution $P(z_{\rm phot}|z_{\rm true})$ in order to assign $z_{\rm phot}$ to each mock galaxy. As detailed in \mocks, we find that the sum of a normal and a normal-skewed distribution works well to reproduce our estimates of $P(z_{\rm phot}|z_{\rm true})$ for the DES Y1 data. This method allows us to accurately model the correspondence between the observed redfshift $z_{\rm phot}$ and the true redshift $z_{\rm true}$, and its effect on the observed clustering. However, small differences remain between the normalized (to integrate to 1) redshift distribution, $\phi(z)$, for the mocks in each redshift bin and that we estimate for the data. Thus, we will use the $\phi(z)$ specific to the mocks for their BAO template.

Galaxies are added to the mocks using a hybrid Halo Occupation Distribution/Halo Abundance Matching model with three parameters. These are each allowed to evolve with redshift in order to account for bias evolution and selection effects. The amplitude of the clustering in the DES Y1 data is reproduced within approximately 1-$\sigma$ in eight $z_{\rm photo}$ bins with $\Delta z_{\rm photo} = 0.05$ in the range $0.6<z_{\rm phot}<1.0$.

Details of the modelling and validation of the mocks can be found in \mocks. Here we use these mock samples to validate our methodology and estimate our covariance matrix, as described in the following section. Other types of mocks have been used in other DES analyses (e.g., \citealt{MacCrann18}), however the Halogen mocks are the only version of DES mocks that have more than 18 realizations, in fact 1800, while also spanning the full Y1 footprint with sufficient resolution to populate halos with the typical galaxies of the BAO sample. Having such a large number of mock samples reduces the noise in the derived covariance matrices and is crucial for identifying the proper procedures for dealing with the particularities of the DES Y1 results.

\section{Analysis}
\label{sec:analysis}
\subsection{Measuring Clustering}

We perform clustering analysis using both the angular correlation function, $w(\theta)$, and the angular power spectrum measured in spherical harmonics, $C_{\ell}$. We also measure the projected comoving separation correlation function, $\xi(s_{\perp},s_{\parallel})$, where $s_{\perp}$ and $s_{\parallel}$ are the apparent transverse and radial separations. Flat sky approximations are never used for the determination of angular separations.

\subsubsection{Angular Clustering}
In order to calculate $w(\theta)$, we create a uniform random sample within the mask defined in Section \ref{sec:benchred} with size 40\ times that of our data sample. We down-sample these randoms given the fractional coverage of the pixel (always $>0.8$ given our mask threshold) to produce the final random sample. Given the random sample, we use the \cite{LS} estimator
\begin{equation}
w(\theta) = \frac{DD(\theta)-2DR(\theta)+RR(\theta)}{RR(\theta)},
\end{equation}
where, e.g., $DD(\theta)$ is the normalized number of pairs of galaxies with angular separation
$\theta\pm\Delta\theta$, with $\Delta\theta$ being half the bin size, and all pair-counts are normalized based on the total size of each sample. We bin paircounts at a bin size of 0.15 degrees, but will combine these to 0.3 degrees for our fiducial bin size (as we will find this to be a more optimal data compression in Section \ref{sec:mocktest}). We will use both binnings as a test of robustness. We will use $0.5 < \theta <5$ degrees for our fiducial scale cuts, yielding 15 $\theta$ bins per redshift bin and thus use 60 total $w(\theta)$ measurement bins. DES-BAO-$\theta$-METHOD demonstrates that adding the results of cross-correlations between redshift bins offers minimal improvement, even when ignoring the degree to which including such measurements would increase the size of the covariance matrix. Thus, we do not consider any such cross-correlations.

The details of the $C_{\ell}$ calculation are presented in \methodcl. They are measured from the decomposition into spherical harmonics of the 
projected 2-dimensional galaxy overdensity $\delta_{\rm gal}$ 
in a given redshift bin. To do so, we use the {\sc anafast} code contained in {\sc HEALPiX} \citep{healpix}. We then use the pseudo-$C_\ell$ method given by \citep{Hivon:2001jp} in order to correct for the effect of the masked sky. In our measurements we use bins of $\Delta \ell = 15$ in the range $30 < \ell < 330$. This $\ell_{\rm max}$
corresponds to a minimum angular scale of $\theta_{min} \approx 0.5^{\circ}$. We thus use 20 $\ell$ bins per redshift bin and 80 total $C_{\ell}$ measurement bins.

\subsubsection{Projected Physical Separation Correlation Function}
We convert our galaxy sample into a three dimensional map in `photometric redshift space' by converting angles and redshifts to physical distances. In this way, we are treating the redshift from the photometric redshift estimate like redshifts are used for calculating clustering statistics for a spectroscopic survey. For the corresponding random sample, we use the same angular coordinates of the randoms in the $w(\theta)$ measurements and assign redshifts to the randoms by randomly selecting redshifts from individual galaxies in our galaxy catalog. We apply a redshift dependent weight, $w_{\rm FKP}(z)$, based on the number density, galaxy bias (determined by interpolating the results in \mocks), and redshift uncertainty as a function of redshift, based on the form derived in \methodXi 
\begin{equation}
w_{\rm FKP}(z) = \frac{b(z)D(z)}{1+n_{\rm eff}(z)P_{\rm lin}(k_{\rm eff},z=0)b^2(z)D^2(z)},
\label{eq:xiFKP}
\end{equation}
where $n_{\rm eff}(z)$ is the effective number density accounting for the redshift uncertainty (using the equations and methodology described in \methodXi), and $k_{\rm eff}$ is the $k$ scale given the greatest weight in Fisher matrix forecasts of the BAO signal, accounting for all of the relevant sample properties. 

We calculated normalized paircounts of galaxies and randoms in bins of 1 $h^{-1}$Mpc along $s_{\perp}$ and $s_{||}$. Calculating the paircounts with this narrow bin size provides the flexiblity to test many different binning schemes. Here, we will combine the paircounts into a bin size of 12$h^{-1}$Mpc for our fiducial measurements, but also present tests with many other bin sizes. Additionally, our final results will combine results across shifts in the center of the bin. This procedure mirrors that used in recent BAO studies \citep{alph,Ata17}. Again, we use a version of the \cite{LS} estimator,  
\begin{equation}
\xi_{\rm phot}(s_{\perp},s_{||}) =\frac{DD(s_{\perp},s_{||})-2DR(s_{\perp},s_{||})+RR(s_{\perp},s_{||})}{RR(s_{\perp},s_{||})}, 
\end{equation}
where $D$ represents the galaxy sample and $R$ represents the uniform random sample that simulates the selection function of the galaxies. $DD(s_{\perp},s_{||})$ thus represents the normalized number of pairs of galaxies with separation $s_{\perp}$ and $s_{||}$. See \methodXi ~for further details. Our fiducial choice employs 14 measurement bins in the range $30 < s_{\perp} < 200 h^{-1}$Mpc. The nature of the measurement means that calculating the clustering over a large redshift window does not dilute the signal like it does for the angular clustering measurements. Therefore, we choose to use the full redshift range when calculating $\xi$. Thus, the size of the data vector (and thus the covariance matrix) is significantly smaller than that of the angular clustering statistics. The changes in the clustering amplitude and number density as a function of redshift are accounted for by the weights given in Eq.~\ref{eq:xiFKP}. The mocks mimic these changes and thus any effects are captured in our mock analysis.

The statistic used for $\xi$ measurements is
\begin{equation}
\xi(s_{\perp}) =  \int_0^{1}{\rm d}\mu_{\rm obs} W(\mu_{\rm obs})\xi_{\rm{phot}}(s_{\perp},s_{||}),
\end{equation}
where the window function $W(\mu_{\rm obs})$ is normalized such that $\int_0^1 {\rm d}\mu_{\rm obs} W(\mu_{\rm obs}) = 1$. $\mu_{\rm obs} = s_{||}/\sqrt{s^2_{\perp}+s^2_{||}}$ is the observed cosine of the angle to the line of sight. We have simply used the data with $\mu_{\rm obs} < 0.8$ and adjusted the normalization to compensate, i.e., our $W(\mu)$ is a step function that is 1 for $\mu < 0.8$ and 0 for $\mu > 0.8$. Once more, this matches the approach advocated in \methodXi, where it was found that the BAO signal to noise and the ability to model it degrades considerably for $\mu > 0.8$. 

\subsection{Covariance and Parameter Inference}
In order to estimate the covariance matrix for our clustering estimates, we use a large number of mock samples, described in Section \ref{sec:mocks}. We have 1800 realizations, so the correlation between data vector, $X$, elements $i$ and $j$ is
\begin{equation}
C_{i,j} = \frac{1}{1799}\sum_{k=1}^{k=1800}(X^k_i-\langle X_i\rangle)(X^k_j-\langle X_j\rangle).
\label{eq:cov}
\end{equation} 
For the angular clustering measurements, the full data vector includes multiple redshift bins and the covariance matrix thus includes terms for the covariance of the clustering between different redshift bins.

For angular clustering statistics, we will compare against results obtained from analytic estimates of the covariance matrix. These estimates assume the statistics are that of a Gaussian field. For $w(\theta)$ this is obtained after transforming to configuration space the following expression,
\begin{equation}
\sigma^2(C_{\ell}) = \frac{2\ell + 1}{f_{\rm sky}(4\pi)^2} \left(C_{\ell}+1/\bar{n}\right)^2.
\end{equation}
Full details are given in section 2.2 of \methodACF, but notably the effect of the survey mask is not included beyond the $f_{\rm sky}$ factor. We denote this as the `Gaussian' covariance matrix. For $C_{\ell}$, the full details are given in \methodcl, but in harmonic space the effect of the shape of the mask is included in the analytic estimate.

The effect of the mask on $w(\theta)$ estimates is studied in \mocks. Two different estimates of the $w(\theta)$ covariance matrix determined using mocks are compared to the analytically estimated Gaussian covariance matrix. One matrix is constructed from mocks with the mask for the DES Y1 BAO sample applied and a separate one is constructed by applying a square mask that has the same area. The results are shown in their figure 12. The diagonal elements of the analytic estimate agree well with the mock estimate using the square mask. The disagreement is worst at small scales for the $0.6 < z < 0.7$ data, which is expected given that this redshift bin has the greatest number density and small scales are expected to have most non-Gaussian influence. When applying the DES Y1 mask to the mocks, significant disagreement is observed at all scales. The mask is also shown to have a significant impact on the off-diagonal component of the covariance matrix. One concludes that the main need to use the mocks to create the $w(\theta)$ covariance matrix is due to the mask. 

\methodACF~ further investigate the connection between the mock and Gaussian covariance matrices. In one test, they determined that the eigenmodes of the $w(\theta)$ covariance matrix determined using 200 mock realizations can be combined with a Gaussian covariance matrix to produce a covariance matrix matching that produced using the full 1800 mock sample. They also showed that an accurate covariance matrix can be produced via
\begin{equation}
C_{\rm new} = C_{\rm old Mock} - C_{\rm old Gauss} + C_{\rm new Gauss},
\label{eq:newcov}
\end{equation}
where `old' denotes the covariance matrix constructed via mocks assuming some old set of parameters that are used in the construction of the Gaussian covariance matrix and `new' denotes some new set of parameters we desire to have the covariance matrix for. We will use Eq. \ref{eq:newcov} in order to test altering the cosmology assumed when constructing the covariance matrix.

\methodcl \, studied the covariance matrix for the $C_{\ell}$ measurements. They applied an analytic method that accounts for the mask, which introduces non-diagonal elements in the covariance, as well as the number density and the level of clustering. At high $\ell$, it was observed that the mock $C_{\ell}$ have lower amplitude than observed for the DES data. The analytic covariance matrix was thus constructed using analytic $C_{\ell}$ matching the observed level of clustering at all scales. This analytic covariance produced a moderate improvement in the recovered $\chi^2$/dof for the best-fit model (85.8/63 compared to 93.7/63). We note that, in contrast to the $C_{\ell}$ results, the mocks are in excellent agreement with the configuration space measurements obtained from the data and we therefore trust the use of mock covariance matrix for such statistics.

The cosmology used to produce the mocks is significantly different from that preferred by current data. The difference in the expected BAO scale when comparing our fiducial cosmology and that preferred by \cite{Planck2015} is 4 per cent, which is a close match to our expected precision. The agreement presented in \mocks~ demonstrates that the clustering of the DES Y1 BAO sample would be unlikely to rule out our fiducial cosmology. Previous studies (e.g., \citealt{Lab12,Taylor13,Morr13,WhitePad15}) on the cosmological dependence of the covariance matrix generally conclude one should be most worried when the data being tested would reject the cosmology assumed to construct the covariance matrix. We test for any impact on our results due to the assumed cosmology explicitly, using Eq. \ref{eq:newcov} to produce a covariance matrix at the \cite{Planck2015} cosmology, in Section \ref{sec:results}.

The impact of choices used in the construction of covariance matrices used for SDSS-III galaxy BAO measurements was studied as part of \cite{Vargas18}. Two sets of mocks were used, using separate approximate methods, as was an analytic approach. They found no significant expected shift in the BAO measurement when using the covariance matrix from one method to measure the BAO on either set of mocks. However, \cite{Vargas18} did find 10 per cent level variation in the recovered size of the uncertainty depending on the covariance matrix that was used. We therefore expect a similar level of uncertainty on our uncertainty determination. We use the covariance matrix determined from our 1800 mock realizations, given by Eq. \ref{eq:cov}, throughout unless otherwise noted.

When using mocks to estimate covariance matrices, we must account for the noise imparted due to the fact we use a finite set of realizations. This noise introduces biases into the inverse covariance matrix. Thus, corrections must be applied to the $\chi^2$ values, the width of the likelihood distribution, and the standard deviation of any parameter determined from the same set of mocks used to define the covariance matrix. These factors are defined in \cite{Hartlap07,Dod13,Per14}. Given that we use 1800 mocks, these factors are at most 3.6 per cent.  

We use the standard $\chi^2$ analysis to quantify the level of agreement between between data and model vectors and to determine the likelihood of parameter values. Given a covariance matrix, ${\sf C}$, representing covariance of the elements of a data vector, and the difference $D$ between a data vector and model data vector, the $\chi^2$ is given by
\begin{equation}
\chi^2 = D{\sf C}^{-1}D^{T}.
\end{equation}
The likelihood, ${\cal L}$, of a given parameter, $p$, is then
\begin{equation}
{\cal L}(p) \propto e^{-\chi^2(p)/2}.
\end{equation}

\subsection{Determining the BAO Scale}

In order to extract the BAO scale from each clustering statistic, we use a template-based method. This approach was used in \cite{Seo12,Xu12,alph,Ross16}. The template is derived from a linear power spectrum, $P_{\rm lin}(k)$, with `damped' BAO modeled using a parameter $\Sigma_{\rm nl}$ (defined below) that accounts for the smearing of the BAO feature due to non-linear structure growth. We first obtain $P_{\rm lin}(k)$ from {\sc Camb}\footnote{camb.info} \citep{camb} and fit for the smooth `no-wiggle'\footnote{Models using only this component will be labeled `noBAO' in plots.} $P_{\rm nw}(k)$ via the \cite{EH98} fitting formulae with a running spectral index. We account for redshift-space distortions (RSD) and non-linear effects via
\begin{equation}
P(k,\mu) = (1+\mu^2\beta)^2\left((P_{\rm lin}-P_{\rm nw})e^{-k^2\Sigma_{\rm nl}^2}+P_{\rm nw}\right),
\label{eq:pkmu}
\end{equation}
where $\mu = {\rm cos}(\theta_{\rm LOS})$ = $k_{||}/k$,
\noindent and $\beta\equiv f/b$. This factor is set based on the galaxy bias, $b$, and effective redshift of the sample we are modeling, with $f$ defined as the logarithmic derivative of the growth factor with respect to the scale factor. The factor $(1+\beta\mu^2)^2$ is the `Kaiser boost' \citep{Kaiser87}, which accounts for linear-theory RSD. The BAO `damping' factor is
\begin{equation}
\Sigma^2_{\rm nl} = (1-\mu^2)\Sigma^2_{\perp}/2+\mu^2\Sigma^2_{||}/2.
\label{eq:damp}
\end{equation}
Given that we have little sensitivity to the line of sight, we will only test varying $\Sigma_{\rm nl}$ (as opposed to its transverse and line of sight components separately), i.e., we use a $\mu$ independent $\Sigma_{\rm nl}$ in Eq. \ref{eq:pkmu}.

Each of $\xi(s_{\perp})$, $w(\theta)$, and $C_{\ell}$ require one or both of the combination of Fourier transforming and projecting Eq. \ref{eq:pkmu} over redshift distributions or uncertainties, in order to obtain the BAO template, $T_{\rm BAO}(x)$, as a function of scale, $x$.\footnote{Here $x$ represents either $r_{\perp}$, $\theta$, or $\ell$ depending on the statistic in question.} For both of the configuration space templates, the anisotropic redshift-space correlation function, $\xi_s(s,\mu)$ is obtained from the Fourier transform of $P(k,\mu)$ defined above. For the angular statistics, we project over the redshift distribution, $\phi(z)$, normalized to integrate to 1. For $w(\theta)$, we have
\begin{equation}
w(\theta) = \int dz_1\int dz_2\phi(z_1)\phi(z_2)\xi_s(s[z_1,z_2,\theta],\mu[z_1,z_2,\theta]).
\end{equation} 
Further details can be found in \methodACF. For harmonic space, we have
\begin{equation}
C_{\ell} = \frac{2}{\pi} \int dkk^2 P(k,\mu=0)\Psi_{\ell}^2,
\end{equation}
with
\begin{equation}
\Psi_{\ell} = \phi(x)j_{\ell}[k\chi(z)],
\end{equation}
where $\chi(z)$ is the comoving distance to $z$ and $\Psi_{\ell}$ is modified to account for RSD as described in \methodcl.

When modeling the projected correlation function $\xi(s_{\perp})$ we follow the formalism of \cite{RossOpt3D}, which strictly speaking is only appropriate for Gaussian photometric redshift errors. Our model for $\xi(s_{\perp})$ is as follows
\begin{equation}
\xi(s_{\perp},\mu) = \int {\rm d}zG(z)\xi(s_{\rm true}[s_{\perp},\mu,z],\mu_{\rm true}[s_{\perp},\mu,z]),
\label{eq:xip}
\end{equation}
where $G(z)$ is a normal distribution of width $\sqrt{2}\sigma_z$ and we used the weighted average of the $\sigma_{68}$ quantities listed in Table \ref{tab:sample}. The $s_{\perp}$ and $\mu$ quantities are those we observe in DES, in the presence of redshift uncertainties, thus requiring the distinction between them and the `true' quantities involved in the projection. See \methodXi~for more details.

Note that unlike our treatment of $\xi(s_{\perp})$, our treatment of the angular correlation function does not assume that the photometric
redshift are Gaussian, and is in fact completely general. This is a primary reason that we adopt our measurement of the angular correlation function as our fiducial analysis. We still explore whether an analysis of the projected correlation function produces consistent results, while possibly reducing the statistical error budget. The impact of the Gaussian photo-$z$ assumption for $\xi(s_{\perp})$ is further discussed in the following sections.

 For each statistic, the BAO scale is obtained through
\begin{equation}
M(x) = BT_{\rm BAO}(x\alpha^{\prime})+A(x),
\end{equation}
where the parameter $\alpha^{\prime}$ rescales the separation to allow a match between the BAO feature in the theory and observation. In configuration space, it is simply $\alpha$ (so, e.g., $x\alpha^{\prime} = \theta\alpha$ for $w(\theta)$, but in harmonic space $\alpha^{\prime} = 1/\alpha$ (so $x\alpha^{\prime} = \ell/\alpha$). Therefore, $\alpha$
parameterizes the BAO measurement (how different the BAO position is in the measurement versus assumed by the template). The parameter $B$ allows the amplitude to change (e.g., due to galaxy bias), and $A(x)$ is a free polynomial meant to account for any differences in between the broadband shape in the data and template. These differences can be due to, e.g., differences between the fiducial and true cosmology or observational systematic effects. Therefore, including the polynomial helps both to isolate BAO scale information and make the measurements robust. Generally, a three term polynomial is used, e.g., for $\xi(s_{\perp})$, $A(s_{\perp}) = a_1 +a_2/s_{\perp}+a_3/s^2_{\perp}$. Similar expressions hold for $A(\theta)$ and $A(\ell)$. Details can be found in \methodACF~and \methodcl.

 For $w(\theta)$ we determine $\Sigma_{\rm nl}$ by fitting to the mean $w(\theta)$ of the mocks. We have fitted to each redshift bin individually. We find that a constant damping scale of 5.2 $\mpcoh$ offers a good fit to all four redshift bins. Based on the modeling described in \methodACF, we expect to find a value consistent with the transverse damping scale for spectroscopic redshift space. Indeed, our recovered value is close to the value of 5.6$\mpcoh$ one obtains when extrapolating the discussion preceding equation 3 of \cite{SeoEis07} to $z=0.8$ and $\sigma_8=0.8$. See \methodACF ~for more details. Thus, we will use this damping scale when fitting to the $w(\theta)$ mocks and data, though we will demonstrate our results are robust against this choice.  

We repeat the procedure in order to find a best-fit $\Sigma_{\rm nl}$ for $\xi(s_{\perp})$. In principle, we should find the same result as found for $w(\theta)$. 
  ~However, our modeling \methodXi ~assumes Gaussian redshift uncertainties, while the true distributions have significant non-Gaussian tails \mocks. Given that the size of the BAO feature depends strongly on the redshift uncertainty, we might expect that our inaccuracies in the treatment of the redshift uncertainty translates to finding a best-fit $\Sigma_{\rm nl}$ that is greater than the theoretical expectation. Indeed, we find $\Sigma_{\rm nl} = 8h^{-1}$Mpc; just over 50 per cent larger than both the value found for $w(\theta)$ and the theoretical expectation. We thus set $\Sigma_{\rm nl} = 8h^{-1}$Mpc for our fiducial $\xi(s_{\perp})$ model, to also account for the known inaccuracies with respect to modeling the redshift uncertainties. We will explore the sensitivity of the results obtained from the DES Y1 mocks and the data to this choice. We will improve the modeling in future analyses.
 
Each method allows us to obtain the likelihood $L(\alpha)$, which represents our BAO measurement. This can be converted to a likelihood for the angular diameter distance $D_A$ at the effective redshift of our sample, $z_{\rm eff}$, via
\begin{equation}
\alpha = \frac{D_A(z_{\rm eff})r^{\rm fid}_{\rm d}}{D^{\rm fid}_A(z_{\rm eff})r_{\rm d}},
\label{eq:alpha}
\end{equation}
where $r_{\rm d}$ is the sound horizon at the drag epoch (and thus represents the expected location of the BAO feature in co-moving distance units, due to the physics of the early Universe). The superscript $^{\rm fid}$ denotes that the fiducial cosmology was used to determine the value. In this work, $r^{\rm fid}_{\rm d} = 153.44$ Mpc. One can see that Eq. \ref{eq:alpha} can be re-arranged to obtain
\begin{equation}
\frac{D_A(z_{\rm eff})}{r_{\rm d}} = \alpha \frac{D^{\rm fid}_A(z_{\rm eff})}{r^{\rm fid}_{\rm d}}.
\end{equation}

The likelihood we obtain for $\alpha\frac{D_A(z_{\rm eff})}{r_{\rm d}}$ can directly be used to constrain cosmological models. In a flat geometry, $D_A$ is given by
\begin{equation}
D_A(z) = \frac{c}{H_0(1+z)}\int_0^z dz^{\prime}\frac{H_0}{H(z^{\prime})}.
\end{equation}
In our fiducial cosmology, $D_A(0.81) = 1597.2$ Mpc. The fiducial $D_A(0.81)/r_{\rm d}$ is thus 10.41. 

 Notably, we are making an implicit assumption that the relative dependence of $D_A$ on cosmological parameters is constant over the redshift range of our sample ($0.6 < z < 1.0$), as even for the statistics where we bin in redshift ($w(\theta)$ and $C_{\ell}$) we are determining a single $\alpha$ likelihood. This is not a perfect assumption, as, e.g., the relative shift in $D_A$ between our fiducial cosmology and the \cite{Planck2015} cosmology between $z=0.6$ and $z=1.0$ is one per cent. This can be compared to the total shift at the effective redshift $z=0.81$ of 4.2 per cent. In effect, our use of the single $\alpha$ means we are not optimally analyzing the signal. \cite{Zhu15} present methodology for a more optimal redshift-space analysis, though \cite{Zhu18} do not find major improvements over the type of analysis we present when applying the methodology to a quasar sample occupying $0.8 < z < 2.2$. Finally, we note that \cite{Bautista18} use the same redshift range for BAO measurements as we do in this analysis.

Our DES Y1 sample is in a regime with an expected signal to noise, in terms of detection ability, of close to 2. In such a regime, we do not expect Gaussian likelihoods. In general for low signal to noise BAO measurements, the tails of the distribution extend to both large and small values of $\alpha$. See, e.g., \cite{Ross15,Ata17} for recent similar signal to noise BAO measurement and \methodACF~for a detailed investigation of what we expect for DES Y1 $w(\theta)$ measurements. Indeed, we find such tails in our DES Y1 analysis and one consequence is that when using our 1800 mock catalogs, we find that 8 per cent of the realizations lead to no clear detection of the BAO feature. An important consequence of these facts is that it is critically important for any cosmological application of our results to consider the full likelihood. We restrict our analysis to $0.8 < \alpha < 1.2$, equivalent to obtaining the posterior likelihood assuming a flat prior on $\alpha$ in this range. This posterior likelihood will be released as a $\chi^2(\alpha)$ lookup table after this work has been accepted for publication.

In the interest of reporting a meaningful summary statistic, we restricted ourselves to the fraction of mock realizations in which the BAO feature was detected (92 per cent), and calculated the error in $\alpha$ by demanding $\Delta\chi^2=1$ relative to the maximum likelihood point. This is the error that we report throughout. Our approach matches that of \cite{Ross15} and \cite{Ata17}, who faced similar issues. When restricting ourselves to mock galaxy catalogs with a BAO detection, we found that this error corresponds to a 68 per cent confidence region for $w(\theta)$ (see \methodACF ~for details, where alternative approaches are also explored and where the approach we adopt is determined to be the best option) and we apply the same criteria to each clustering statistic we use, as each is fit using the same basic methodology. Thus, while this quantity is not formally a 68 per cent confidence region for our posterior likelihood, we have opted for utilizing this quantity as a summary statistic. In practice, all cosmological inferences from our results will utilize our full posterior likelihood.

\section{Tests on Mocks}
\label{sec:mocktest}

In this section, we report the results of testing our BAO fitting methodology on the 1800 mock realizations. We test fits to both the mean of these mocks and each mock individually. These tests inform how we obtain our final consensus Y1 BAO measurement and how we decide fiducial settings such as bin size and the range of scales considered. We report the results of tests for each clustering statistic we present BAO measurements for. Additional tests for $\xi(s_{\perp})$ and $w(\theta)$ measurements can be found in \methodXi~and \methodACF, with implications beyond the DES Y1 sample. The motivation for fiducial choices for the $w(\theta)$ analysis are described in \methodACF. We divide the section into tests done on the mean of the mocks (giving us one data vector with the signal to noise for 1800 DES Y1) and tests done on each individual mock (providing distributions for the signal to noise we should expect to recover).

\begin{table}
\centering
\caption{ The expected uncertainty for DES Y1 data, assuming a Gaussian likelihood applied to the mean $\xi(s_{\perp})$ obtained from 1800 mock realizations, as a function of the $s_{\perp}$ binning that is used. See text for details. }  
\begin{tabular}{lcc}
\hline
\hline
binning & $\sigma_{\rm G}$  \\
\hline
$0.6 < z < 1.0$:\\
$\Delta s_{\perp} = 5h^{-1}$Mpc & 0.054\\
$\Delta s_{\perp} = 8h^{-1}$Mpc & 0.053\\
$\Delta s_{\perp} = 10h^{-1}$Mpc & 0.052 \\
$\Delta s_{\perp} = 12h^{-1}$Mpc & 0.051\\
$\Delta s_{\perp} = 15h^{-1}$Mpc & 0.052\\
$\Delta s_{\perp} = 20h^{-1}$Mpc & 0.059\\
\hline
\label{tab:xibin}
\end{tabular}
\end{table}

\subsection{Test on mean statistics}
\label{sec:testonmean}
The total number of mock realizations has high signal to noise. In principle, we should divide our covariance matrix by 1800 in order to fit the mean of the mocks. However, we are primarily interested in the uncertainty we should expect for DES Y1, and thus we will quote results obtained either from the nominal covariance matrix for DES Y1 or with the appropriate scaling.

First, we determine the fiducial bin size for the $\xi(s_{\perp})$ analysis by fitting to the mean $\xi(s_{\perp})$ of the 1800 mock realizations. If not for noise from the covariance matrix, using the smallest bin size possible would always maximize the signal to noise. However, the noise in the covariance matrix increases with its number of elements and thus the optimal bin size will be somewhat greater than the size where significant information starts to be lost. The signal to noise for one realization is such that the likelihoods are typically non-Gaussian. As the signal to noise of BAO measurements increases, the likelihoods typically become well-approximated by Gaussians (e.g., compare \citealt{Ross15} to \citealt{alph}). To take advantage of this, we divide the DES Y1 covariance matrix obtained from the mocks by 10 and obtain the likelihood. We then obtain $\alpha$ and $\sigma$ as usual but define a `Gaussian' uncertainty $\sigma_G = \sqrt{10}\sigma$. The results are unchanged if we use a factor of 20 rather than our factor of 10. The results are presented in Table \ref{tab:xibin}. We find that the optimal results are expected for a bin size of 12$h^{-1}$Mpc. This is significantly greater than the optimal bin size typically found for spectroscopic surveys; a potential explanation is that the redshift uncertainty has significantly smeared the BAO, making a narrow bin size less important for recovering the total signal (see figure 1 of \citealt{RossOpt3D}).

\begin{table}
\centering
\caption{ BAO fits to the mean Y1 mocks. The $\alpha$ values suggest how biased our fitting methods are and the $\sigma$ represents something akin to a Fisher matrix prediction for the precision we should achieve on the data. The fiducial analysis choices for $\xi(s_{\perp}$ are $30 \leq s_{\perp} < 200h^{-1}$Mpc and $\Delta s_{\perp} = 12h^{-1}$Mpc. For $w(\theta)$, they are $0.5 < \theta < 5$ degrees and $\Delta \theta = 0.3$ degrees.}  
\begin{tabular}{lc}
\hline
\hline
case & $\alpha$    \\
\hline
$0.6 < z < 1.0$:\\
$w(\theta)$ & $1.003\pm0.055$   \\
$w(\theta)$, $\theta_{\rm min} =1$ deg& $1.003 \pm 0.055 $    \\
$w(\theta)$, $\Delta \theta =0.15$ deg& $1.004\pm0.057$   \\
$C_{\ell}$ & $1.009 \pm 0.056$\\
$\xi$ & $1.007 \pm 0.052$   \\
$\xi$, $s_{\perp, \rm min}=50h^{-1}$Mpc &$1.008 \pm 0.052 $   \\
$\xi$, $\Sigma_{\rm nl} = 4h^{-1}$Mpc & $1.005\pm0.045$ \\
$\xi$, $\Sigma_{\rm nl} = 12h^{-1}$Mpc & $1.009\pm0.065$ \\
\hline
\label{tab:baomean}
\end{tabular}
\end{table}

Table \ref{tab:baomean} displays results for fits to the mean of the 1800 mocks, using the DES Y1 covariance matrix. For our fiducial analysis choices, we expect an uncertainty of just greater than 5 per cent.  
We also see that choosing $s_{\perp, \rm min} = 30 h^{-1}$Mpc opposed to $50 h^{-1}$Mpc improves the results both in terms of the bias in $\alpha$ and the recovered uncertainty. For $w(\theta)$ using a larger bin size of 0.3 degrees improves the results compared to 0.15 degrees. This is due to the fact that the number of elements in the covariance matrix is reduced from $120^2$ to $60^2$, significantly reducing the required correction factors. \methodACF ~reports further tests of the bin size, suggesting no significant improvement is to be achieved compared to the fiducial 0.3 degree bin size. We further see that we expect to recover slightly smaller uncertainties from $\xi$ compared to $w(\theta)$ or $C_{\ell}$, but this is at most a 5 per cent difference. Further tests of the $C_{\ell}$ are detailed in \cite{clbao}.

The $\alpha$ obtained from the mean of the mocks is biased high for all three methods we have tested. For the uncertainty expected from a single Y1 realization, it is a 0.06$\sigma$ (0.003) bias for $w(\theta)$, 0.13$\sigma$ (0.007) for $\xi$, and 0.16$\sigma$ (0.009) for $C_{\ell}$. This is small enough to not be a significant concern for the Y1 signal to noise. However, given that this is the mean of 1800 mocks, the significance of the detection of a bias is 6.8$\sigma$ for $C_{\ell}$, 5.6$\sigma$ for $\xi$, and only 2.3$\sigma$ for $w(\theta)$. This suggests it is a true bias that will need to be addressed as the signal to noise increases for future data samples. We will use the $w(\theta)$ results for our DES Y1 measurement, where the bias is only of marginal significance. As detailed in, e.g., both \cite{Crocce08} and \cite{PadWhite09}, a small positive bias is expected from non-linear structure growth, which could explain $\sim$0.003 worth of the bias and thus fully account for the $w(\theta)$ results.

\subsection{Tests on individual mocks}

\begin{table}
\centering
\caption{
Statistics for BAO fits on mocks. $\langle\alpha\rangle$ is either the BAO dilation-scale measured from the correlation function averaged over all of the mocks (denoted `mean'), or the mean of the set of dilation-scales recovered from mocks with $>1\sigma$ BAO detections. $\langle\sigma\rangle$ is the same for the uncertainty obtained from $\Delta\chi^2=1$ region. $S$ is the standard deviation of the $\alpha$ recovered from the mock realizations with $>1\sigma$ BAO detections and $f(N_{\rm det})$ is the fraction of realizations satisfying the given condition. }
\begin{tabular}{lcccc}
\hline
\hline
case & $\langle\alpha\rangle$ & $\langle\sigma\rangle$ & $S_{\alpha}$ & $f(N_{\rm det})$ \\
\hline
$0.6 < z < 1.0$: \\
$\xi+w$ & 1.004 & 0.050 & 0.050 & 0.917   \\
$w+C_{\ell}$ & 1.006 & 0.055 & 0.052 & 0.901 \\
$w(\theta)$ & 1.001 & 0.051 & 0.054 & 0.898   \\
$w(\theta)$, $\Delta \theta =0.15$ deg  & 1.001 & 0.054 & 0.055 & 0.907 \\
$w(\theta)$, $\theta_{ \rm{min}} =1$ deg & 1.002 & 0.051 & 0.053 & 0.898 \\
$C_{\ell}$ & 1.007 & 0.058 & 0.053 & 0.864\\
$\xi$ (bins combined) & $ 1.004 $ & $ 0.048 $ & $ 0.050 $ & $ 0.916 $ \\
$\xi,$ $+0h^{-1}$ Mpc & $ 1.004 $ & $ 0.048 $ & $ 0.050 $ & $ 0.916 $ \\
$\xi,$ $+3h^{-1}$ Mpc & $ 1.004 $ & $ 0.048 $ & $ 0.051 $ & $ 0.916 $ \\
$\xi,$ $+6h^{-1}$ Mpc & $ 1.005 $ & $ 0.048 $ & $ 0.050 $ & $ 0.916 $ \\
$\xi,$ $+9h^{-1}$ Mpc & $ 1.005 $ & $ 0.048 $ & $ 0.050 $ & $ 0.921 $ \\
$\xi$, $s_{\perp \rm{min}} = 50h^{-1}$ Mpc  &$ 1.005 $ & $ 0.049 $ & $ 0.050 $ & $ 0.913 $ \\
$\xi$, $\Delta s_{\perp} = 5h^{-1}$ Mpc & 1.005 & 0.050 & 0.051 & 0.918\\
$\xi$, $\Delta s_{\perp} = 10h^{-1}$ Mpc & 1.005 & 0.049 & 0.050 & 0.916\\
$\xi$, $\Delta s_{\perp} = 15h^{-1}$ Mpc & 1.004 & 0.048 & 0.051 & 0.911\\
\hline
\label{tab:baomock}
\end{tabular}
\end{table}

\begin{figure}
\includegraphics[width=84mm]{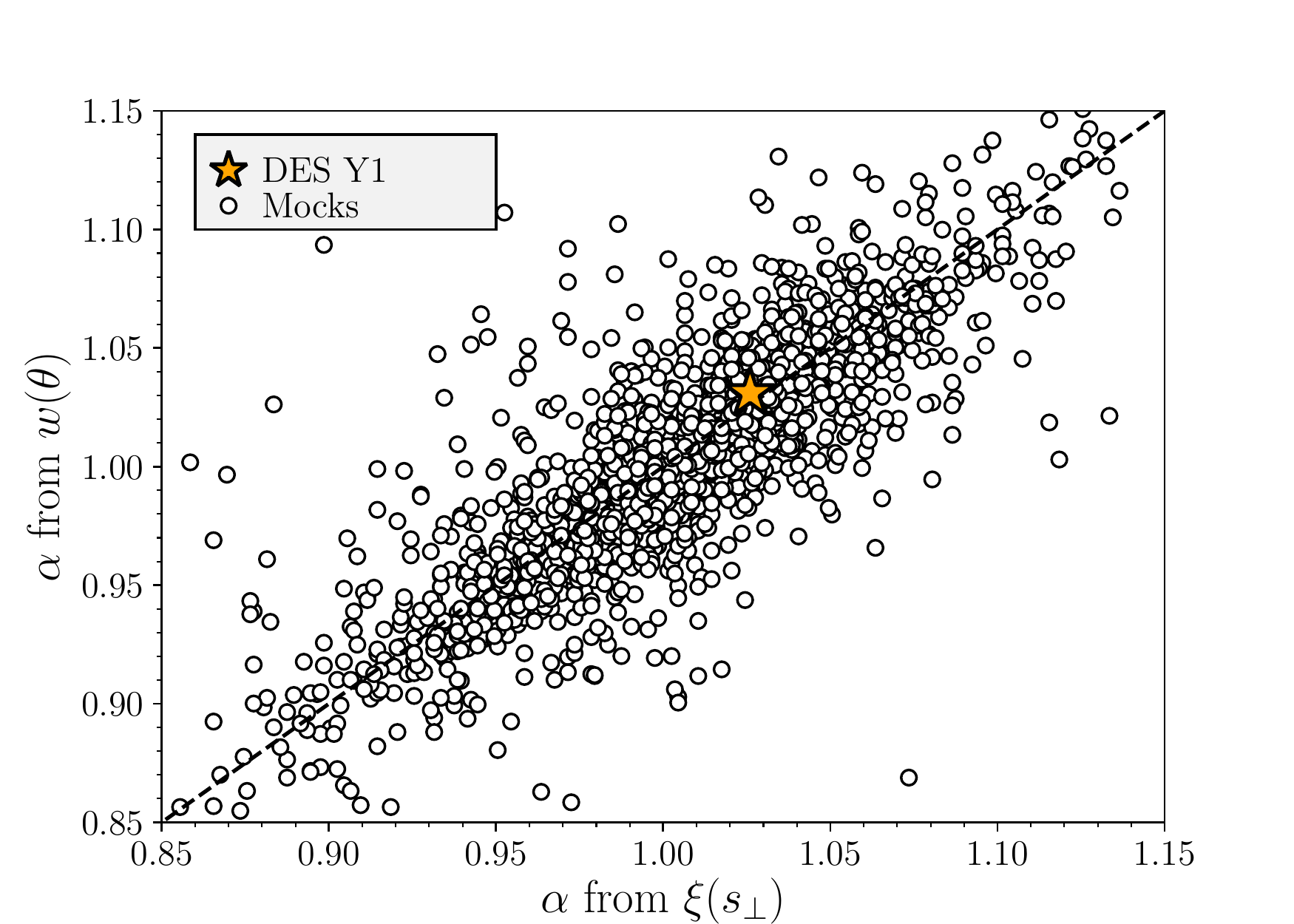}
\includegraphics[width=84mm]{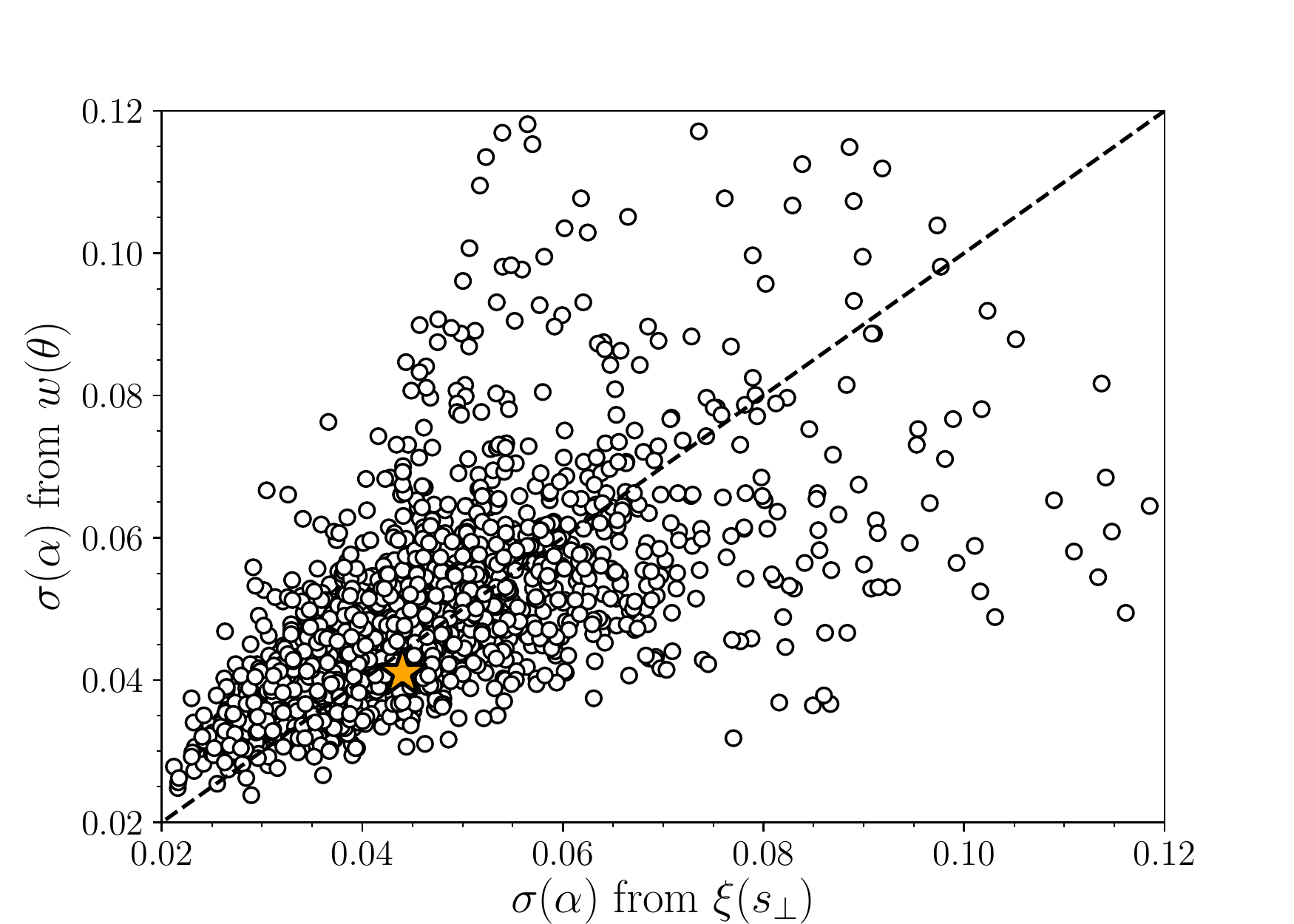}
  \caption{A comparison of $w(\theta)$ and $\xi(s_{\perp})$ BAO fit parameter $\alpha$ and its uncertainty performed on mock realizations (white circles) and the DESY1 data (stars). The mock realizations are for $0.6 < z < 1.0$. The uncertainty, $\sigma$, is obtained from the $\Delta \chi^2=1$ definition (see text). 
  }
  \label{fig:BAOwxi}
\end{figure}

\begin{figure}
\includegraphics[width=84mm]{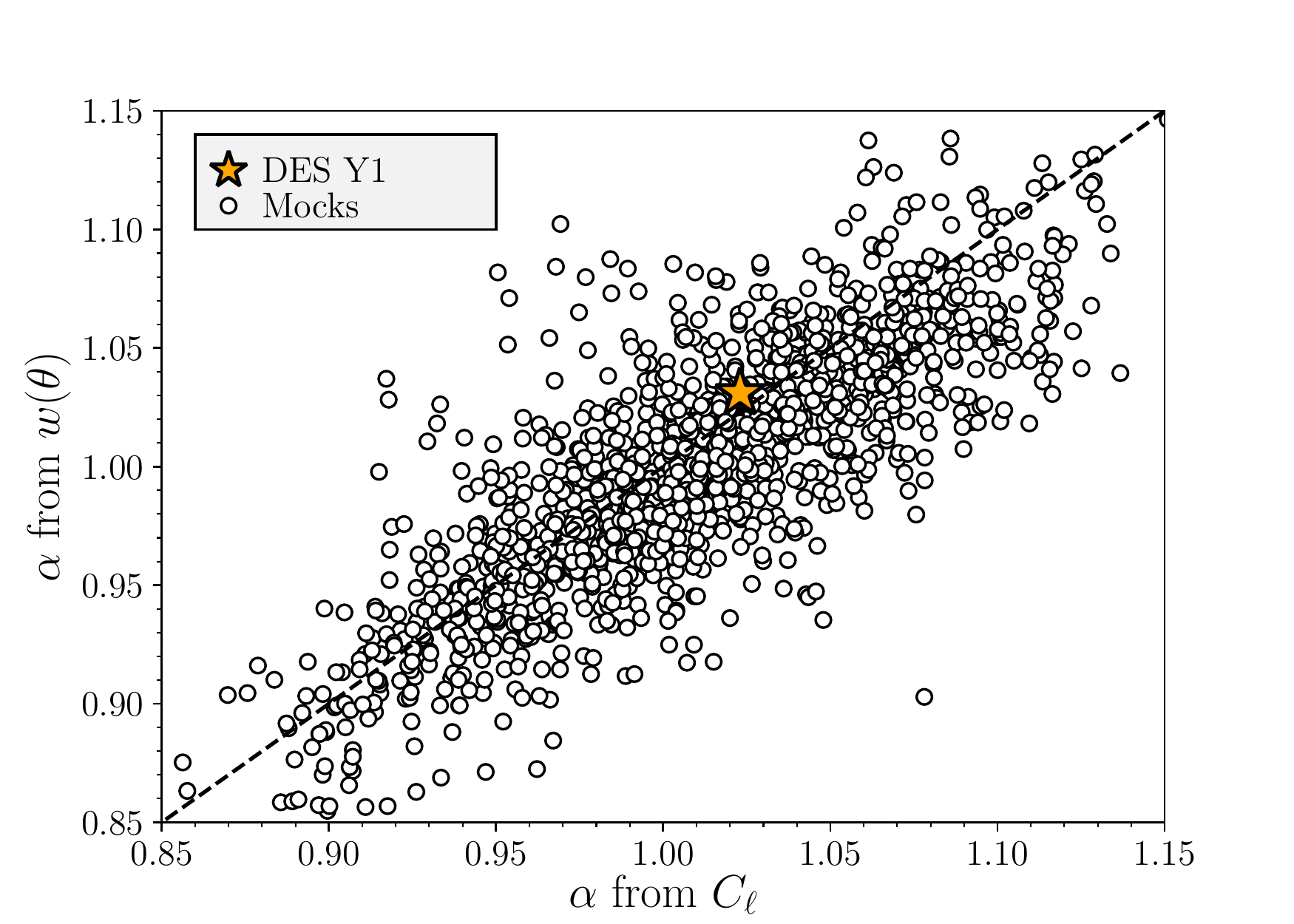}
\includegraphics[width=84mm]{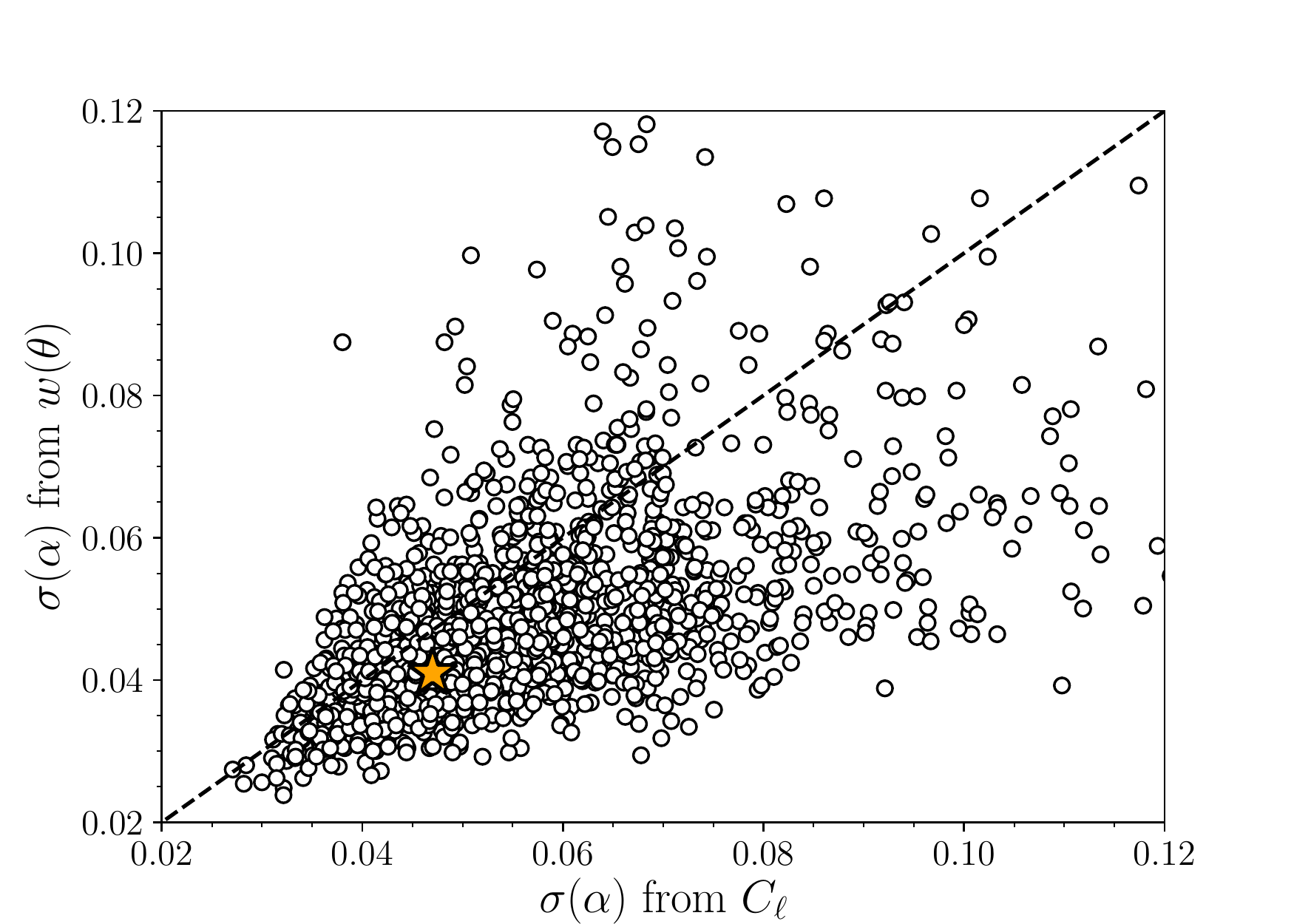}
  \caption{Same as Fig. \ref{fig:BAOwxi}, but the $C_{\ell}$ results replace $\xi(s_{\perp})$. 
  }
  \label{fig:BAOwcl}
\end{figure}

Results obtained from fitting each individual mock realization are shown in Table \ref{tab:baomock}. We denote the mean and standard deviation of any measured quantity $x$ across realizations using $\langle x\rangle$ and $S_x$. Results are shown for cases where there is a $\Delta \chi^2 =1$ region within $0.8 < \alpha < 1.2$; these are referred to as `detections' (and those mocks where this is not the case are `non-detections'). For $\xi(s_{\perp})$ just over 91 per cent of the mocks yield a detection while for $w(\theta)$ it is just less than 90 per cent. The results are generally consistent with the tests on the mean of the mocks. We learn that the standard deviation and mean uncertainties are matched to within 4 per cent. The mean uncertainties are generally slightly smaller than the standard deviations, reflecting the fact that the likelihoods have non-Gaussian tails and we are using $\Delta \chi^2=1$ to quote the uncertainty. 

For all three methods, the biases in $\alpha$ have decreased slightly, though this is likely due to our detection criteria within $0.8 < \alpha < 1.2$ (since it is symmetric around $\alpha =1$ instead of $\alpha \sim 1.005$). The uncertainties for $w(\theta)$ are only 5 per cent greater than for $\xi$, and the $C_{\ell}$ results are somewhat less precise than the $w(\theta)$ results. In configuration space, varying the bin size or minimum scale does not reveal any large changes in the results. Further tests are performed on the $C_{\ell}$ measurements in \methodcl . In particular, the mocks are used to determine the optimal range in $\ell$, the bin size in $\ell$, and the number and type of polynomial broadband terms to use. The results presented here show their optimized choices.


For $\xi(s_{\perp})$, we also vary the center of the bin, in steps of $3h^{-1}$Mpc, and combine the results by taking the mean of the resulting four likelihoods. This process is similar to that of \cite{Ross15} and \cite{Ata17}, where it was found such a procedure provides small improvements in the accuracy of both the recovered $\alpha$ and its uncertainty. We find that this process has a small effect on the results. The standard deviation is not improved at the level reported in Table \ref{tab:baomock}, but comparing the combined result to the $+0h^{-1}$Mpc result, there is a one per cent improvement in the standard deviation for the combined results. The biggest change from combining the likelihoods is that there is somewhat less dispersion in the uncertainty recovered from the likelihood. In the $+0h^{-1}$Mpc case, the standard deviation of the uncertainties is 0.018, while after combining it is reduced to 0.017. We also determine the standard deviation of the scatter, per mock, for the results in each of the four bin centers.
We find 0.004 (so this is the level of difference we expect to find when repeating these tests on the DES Y1 data).

Fig. \ref{fig:BAOwxi} compares the results of BAO fits to the mocks for $\xi$ and $w(\theta)$ using white circles. The results are shown only for realizations that have a detection for both statistics, which is 1565 realizations (87 per cent). Stars represent the results for the DES Y1 data and are discussed in Section \ref{sec:results}. The bottom panel displays the results for the value of $\alpha$. As expected, the two results are correlated, though there is significant scatter. The correlation factor is 0.81, while for these realizations the standard deviation in the $\xi$ results is 0.048 and it is 0.051 for $w(\theta)$. Taking their mean, the standard deviation is reduced to 0.047; this suggests some small gain is possible from combining the measurements. The top panel displays the results for the recovered uncertainty. $\xi$ recovers a lower uncertainty on average, but there is a large amount of scatter. We test our results when taking the mean of the $\xi(s_{\perp})$ and $w(\theta)$ likelihoods, labeled `$\xi+w$' in Table \ref{tab:baomock}. We find mean uncertainty matching the standard deviation at 0.050 and the highest fraction of `detections'. However, the gain in the precision from the combination is similar to the shift in $\alpha$ (away from the unbiased value of 1), suggesting the gain from combining the results is not worthwhile for the cases where $w(\theta)$ has sufficient signal to noise on its own for a robust measurement.

Fig. \ref{fig:BAOwcl} repeats the comparison, but substitutes the $C_{\ell}$ results for $\xi$. 1502 (83 per cent of) mock realizations have a $\Delta \chi^2 = 1$ bound within $0.8 < z < 1.2$ for both statistics. As expected, the results are strongly correlated in $\alpha$, with a correlation factor of 0.80. The orange star represents the $\alpha$ values recovered for the DES Y1 data; the fact that it lies within the locus of points representing the mock realizations suggests the differences we found in $\alpha$ are typical. The same is true for the recovered uncertainty, where there is a fairly large dispersion but the uncertainty recovered from the $C_{\ell}$ measurements is greater on average. Correspondingly, for this selection of mock realizations the standard deviation in $\alpha$ is slightly greater for the $C_{\ell}$: 0.052 compared to 0.051. Like for the the $\xi+w$ results, we obtain results by using the mean of the $C_{\ell}$ and $w(\theta)$ likelihoods. Similar to the $\xi+w$ case, the number of detections and the standard deviation are improved over the case of using either statistic alone, but the $\langle \alpha \rangle$ has shifted away from 1 to become more biased and this shift greater than the gain in precision. The result once more suggests that combining the results is unlikely to worthwhile for the DES Y1 data. 

The strength of the BAO feature, and thus its signal to noise, in any particular realization of the data can vary. This is clear from the wide range of uncertainties shown in Fig. \ref{fig:BAOwxi}, and is consistent with previous BAO analyses (see, e.g., figure 10 of \citealt{Ata17}). We can use the mocks to determine the extent to which the scatter in the uncertainties recovered from the likelihood are truly representative of the variance in the ability to estimate the BAO parameter $\alpha$. We do so by dividing the mock samples into bins based on the recovered uncertainty and comparing to the standard deviation of $\alpha$ values in each bin, using the mean of the $\xi+w$ likelihoods. Dividing into bins with approximately the same number of mocks in each (to within 30 mocks), the mean uncertainty and standard deviations are $\langle\sigma\rangle , S_{\alpha} = (0.035,0.039), (0.043,0.049), (0.052,0.054), (0.073,0.055)$. For the mock realizations with the highest uncertainty, the scatter in $\alpha$ values is significantly smaller. This is likely due to the fact that the $\alpha$ values must lie within (0.8+$\sigma$,1.2-$\sigma$) in order to be counted as a detection and this therefore decreases their standard deviation. At lower values of uncertainty, there is a clear correlation between the mean recovered uncertainty and the scatter in best-fit $\alpha$. The standard deviations are found to be somewhat larger than the mean uncertainties, likely due to the fact that the likelihoods are non-Gaussian. These results suggest that, generally, we can trust the individual likelihoods (more so than, e.g., taking the mean shape and width of the likelihood of the mocks), especially in the cases with the best apparent signal to noise.

The results of this section can be summarized as follows: The $w(\theta)$ results are the least biased; their bias is only at a 2.3$\sigma$ level, when considering the combined precision of all 1800 mocks, and at least part of the this bias can be explained by the positive bias expected from non-linear structure growth. The $C_{\ell}$ and $\xi$ results are each biased at more than $5\sigma$, based on the combined precision of all of the mocks. The $\xi$ results are the most precise of any method and are the most likely to obtain a detection. Combining either the $\xi$ or $C_{\ell}$ statistics with the $w(\theta)$ results can produce small improvements in the precision at the cost of increasing the bias in the $\alpha$ measurement. We determine this increased bias is unlikely to be worth the gain in precision, but leave any final determinations to be based on analysis of the DES Y1 data.

\section{Results}
\label{sec:results}

\begin{figure}
\includegraphics[width=80mm]{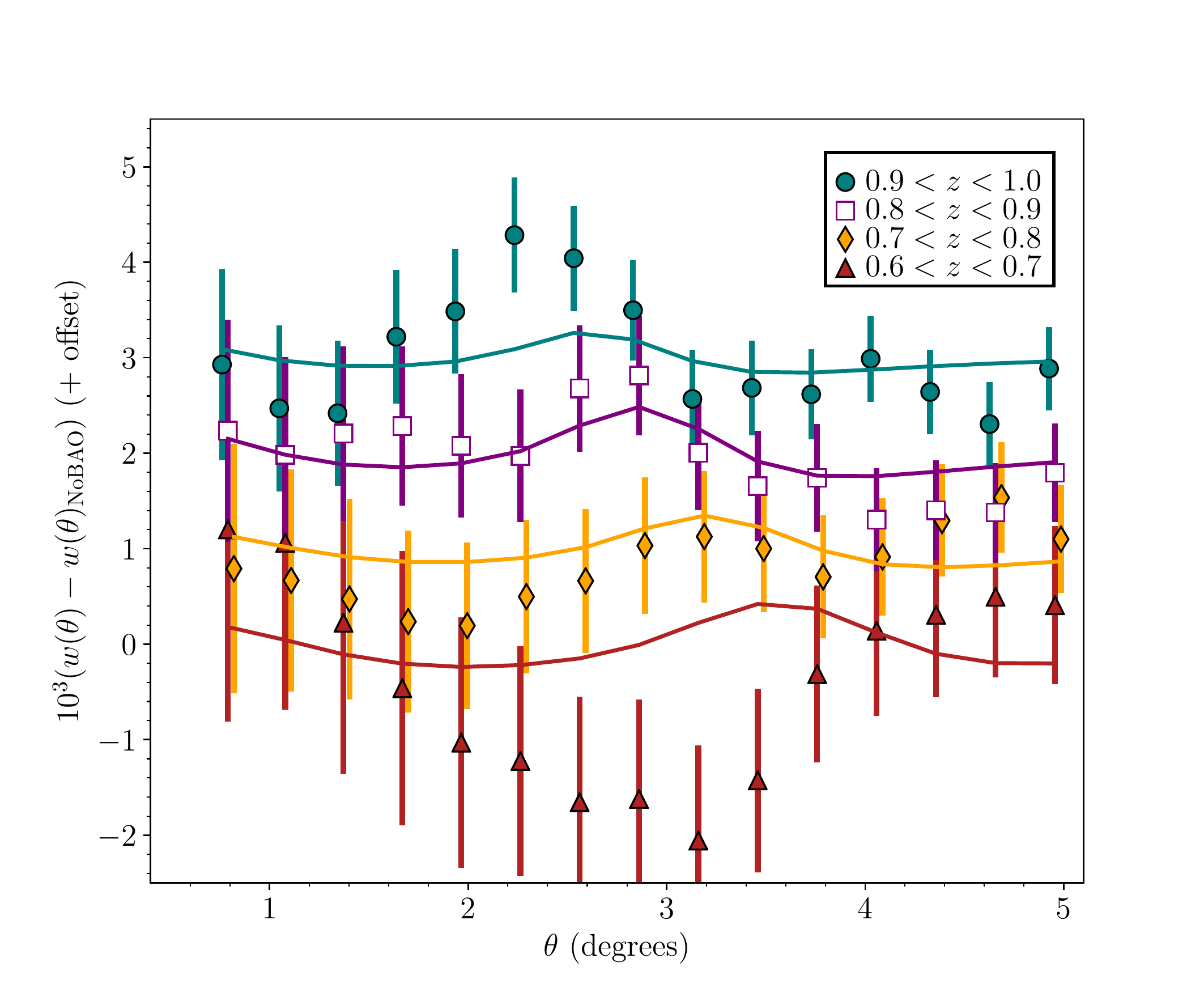}
  \caption{The BAO signal in DES Y1 clustering, observed in the angular auto-correlation, $w(\theta)$ and isolated by subtracting the no BAO component of the best-fit model. The result has been multiplied by $10^3$ and we add vertical offsets of 0, 1, 2, and 3 sequentially with redshift. The $\theta$ values have been shifted by 0.03 for the $0.7 < z < 0.8$ and by -0.03 for the $0.9 < z < 1.0$ redshift bins. The BAO feature moves to lower $\theta$ at higher redshift, as it has the same co-moving physical scale. The signal from these redshift bins is combined, accounting for the covariance between them, in order to provide a 4 per cent angular diameter distance measurement at the effective redshift of the full sample. Neighboring data points are strongly correlated. The total $\chi^2/{\rm dof}$ (including all cross-covariance between redshift bins) is 53/43 and other studies show that, despite its appearance, the $0.6 < z < 0.7$ bin has a $\chi^2/{\rm dof}\sim 1$. }
  \label{fig:wthBAO}
\end{figure}

\begin{figure}
\includegraphics[width=80mm]{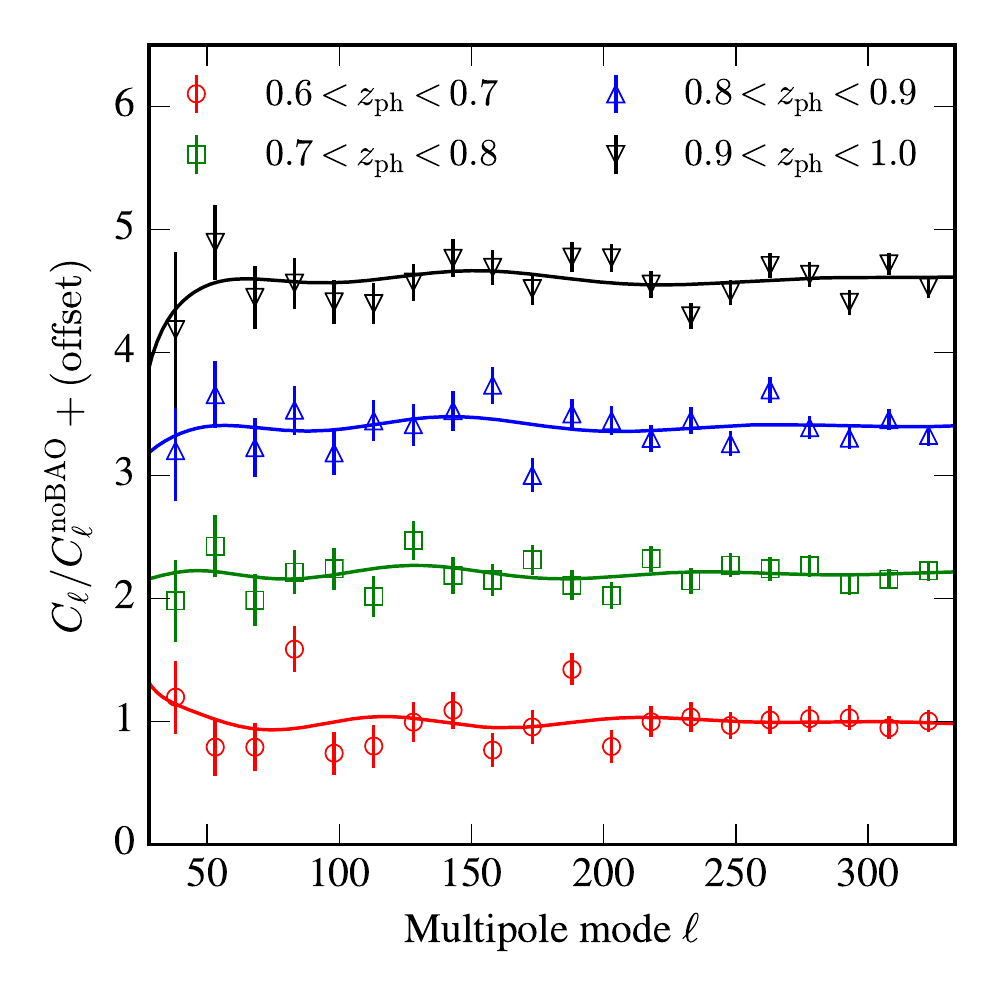}
 \caption{The measured Y1 BAO feature, same as Fig. \ref{fig:wthBAO}, but isolated in spherical harmonic space. From top to bottom, one can see that the BAO feature moves to the left, towards lower $\ell$, reflecting the redshift evolution of a feature of constant co-moving size.}
  \label{fig:clbao}
\end{figure}

\begin{figure}
\includegraphics[width=84mm]{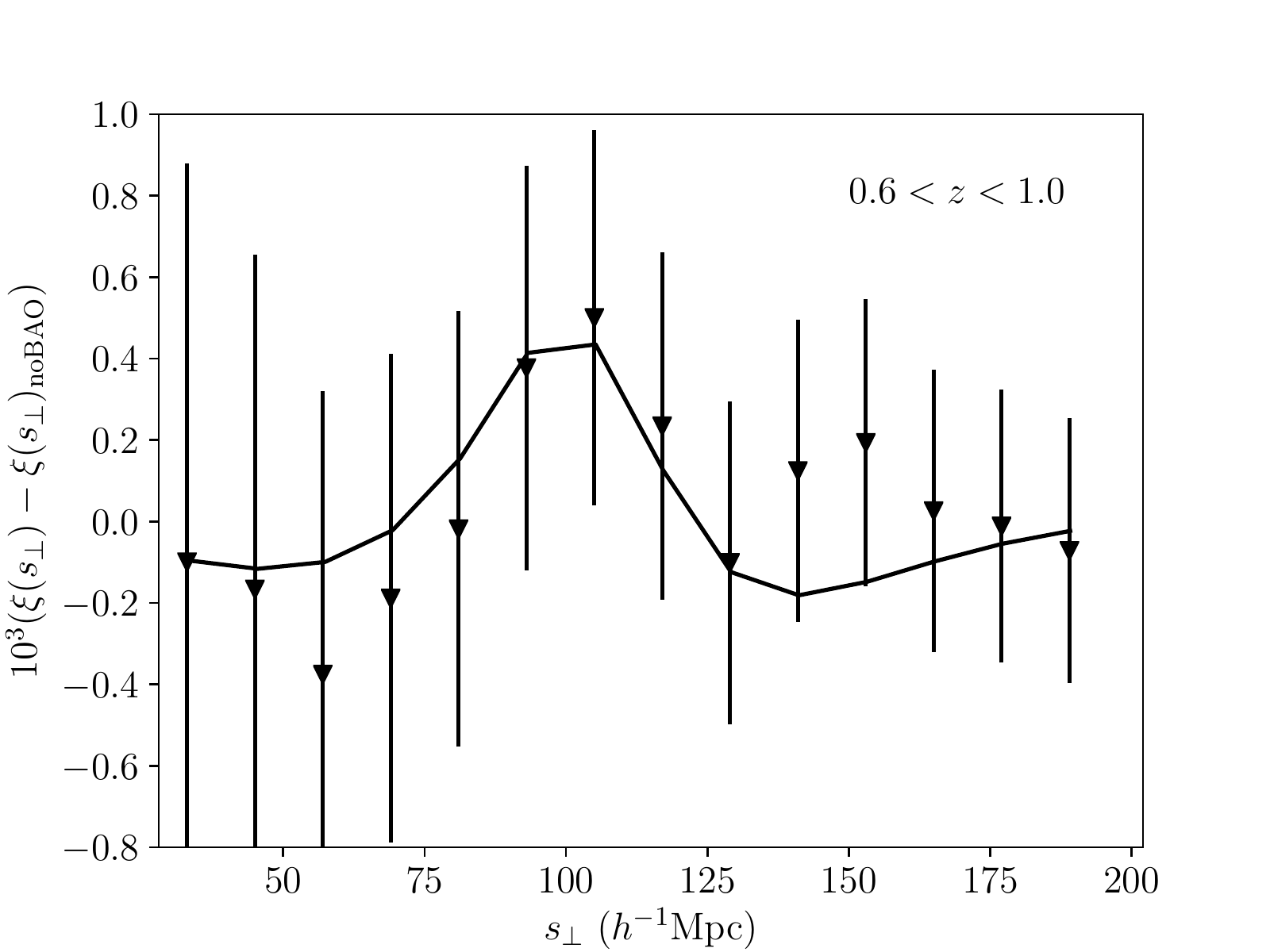}
  \caption{The BAO signal in DES Y1 clustering, observed in the auto-correlation binned in projected physical separation, $\xi(s_{\perp}),$ and isolated by subtracting the no BAO component of the best-fit model. Neighboring data points are strongly correlated.}
  \label{fig:clusBAO}
\end{figure}

Here we focus only on the BAO signal. The validation of the full shape of the clustering signal of the DES Y1 BAO sample is presented in \mocks ~and \sample. \mocks ~shows that both the angular and $\xi(s_{\perp})$ clustering measurements agree well with the clustering in the mock samples. \sample ~shows that the clustering is well-matched to expectations of linear theory in all of our redshift bins and that the galaxy bias evolves from approximately 1.8 to 2.0 within our $0.6 < z < 1.0$ redshift range. \sample ~ also shows that the impact from observational systematics, determined by comparing the clustering with and without the systematics weights, is small. We refer the reader to \sample ~for clustering measurements where the non-BAO information is included. Harmonic space measurements and interpretation are presented in \methodcl. We present the configuration space DES Y1 BAO signal, determined by subtracting the 
`no wiggle' component (see Eq. \ref{eq:pkmu} and surrounding discussion) of the best-fit model (labeled as `noBAO' in figures). We present the DES Y1 measurements of the angular diameter distance to $z=0.81$ in the following subsection and describe the series of robustness tests we apply to the data in Section \ref{sec:robust}.


\begin{table}
\centering
\caption{Results for BAO fits to the Y1 data. The top line quotes our consensus DES Y1 result from $w(\theta)$ in terms of the physical distance ratio $D_A(z=0.81)/r_d$. The other lines report measurements of $\alpha$, which represent the measured shift in $D_A(z=0.81)/r_d$ relative to our fiducial MICE cosmology; e.g., the value expected for Planck $\Lambda$CDM is $\alpha=1.042$. All results assume a flat prior $0.8 < \alpha < 1.2$. Robustness tests against our fiducial analysis settings are reported. These settings include: We use the full $0.6 < z < 1.0$ data set; the binning in $w(\theta)$ is 0.3 degrees and its range is $0.5 < \theta < 5$ degrees; the binning in $\xi$ is $\Delta s_{\perp} =12h^{-1}$Mpc, the range of included bin centres is $30 < s_{\perp} < 200 h^{-1}$Mpc, and the first included bin centre allows pairs with  $27 < s_{\perp} < 39h^{-1}$Mpc. The `bins combined' $\xi$ result is derived from the mean likelihood of the fiducial result and three additional bin centres, shifted in steps of $3h^{-1}$Mpc (and each individual result is denoted below by +/-$x$). `BPZ' denotes that the BPZ photozs were used, as opposed to the fiducial DNF and `$z$ uncal' refers to the case where we use the redshift distribution reported by DNF without any additional calibration for determining the theoretical template. For cases where we alter the assumed $\Sigma_{\rm nl}$ in the template, the units of the quoted values are $h^{-1}$Mpc; the fiducial values are $8h^{-1}$Mpc for $\xi$ and $5.2h^{-1}$Mpc for $w(\theta)$ and $C_{\ell}$.  `Planck' denotes the case where a cosmology consistent with Planck $\Lambda$CDM has been used to calculate paircounts and the BAO template. $A=0$ denotes that no broad-band polynomial was used in the fit, while $a_x$ denotes variations on the terms that were included.}
\begin{tabular}{lcc}
\hline
\hline
{\bf Y1 Measurement} & $D_A/r_{\rm d}$   \\
 $z_{\rm eff}=0.81$ & $10.75\pm 0.43$  \\
\hline
case & $\alpha$  &  $\chi^2$/dof\\
\hline
 $w(\theta)$ [{\bf Y1 choice}] & $ 1.033 \pm 0.041$   & $53/43$ \\
\hline
Robustness tests:\\
$C_{\ell}$  & $1.023\pm 0.047$  & $94/63$\\
$C_{\ell}$  alt cov. & $1.039\pm 0.053$  & $86/63$\\
$w(\theta) $ fiducial & $1.033 \pm 0.041$   & $53/43$ \\
$w(\theta) $ $\Delta \theta = 0.15$ & $ 1.033 \pm 0.045$   & $159/103$ \\
$w(\theta)$ $\theta_{\rm min}=1$ & $ 1.038 \pm 0.038 $     & 50/39\\
$w(\theta)$ Planck$\times$1.042 &  $1.034 \pm 0.043$   & 44/43 \\
$w(\theta)$ BPZ & $1.018 \pm 0.043   $ & 56/43 \\
$w(\theta)$ $z$ uncal & $1.023 \pm 0.040 $ & 52/43 \\
$w(\theta)$ no $w_{\rm sys}$ & $1.028 \pm 0.039  $  & 51/43   \\
$w(\theta)$ $\Sigma_{\rm nl}=2.6$ & $1.028 \pm 0.035 $  & 51/43\\
$w(\theta)$ $\Sigma_{\rm nl}=7.8$ & $1.033 \pm 0.056 $ &  55/43\\
$w(\theta)$  free $\Sigma_{\rm nl}$ & $1.028 \pm 0.033 $  & 51/42\\
$w(\theta)$ $0.7 < z < 1.0$ & $ 1.053 \pm 0.040$   & $37/32$ \\
$w(\theta)$ $A=0$  & $ 1.030 \pm 0.040$   & $59/55$ \\
$w(\theta)$ Gaussian cov.   & $ 1.038 \pm 0.033$   & $88/43$ \\
$ \xi$ (bins combined) & $1.026 \pm 0.044$  & $9/9$ \\
$ \xi$ fiducial binning & $1.031 \pm 0.040$  & $9/9$ \\
$ \xi$ $-3$ & $1.031 \pm 0.045$  & $12/9$ \\
$ \xi$ $+3$ & $1.017 \pm 0.041$  & $8/9$ \\
$ \xi$ $+6$ & $1.025 \pm 0.050$  & $7/8$ \\
$ \xi$ $\Delta s_{\perp} = 5$ & $1.021 \pm 0.041$  & $45/29$ \\
$ \xi$ $\Delta s_{\perp} = 8$ & $1.029 \pm 0.046$ &  $31/16$ \\
$ \xi$ $\Delta s_{\perp} = 10$ & $1.022 \pm 0.037$ &  $16/12$ \\
$ \xi$ $\Delta s_{\perp} = 15$ & $1.012 \pm 0.039$ &  $7.5/6$ \\
$ \xi$ $s_{\perp ,\rm min}=50 $ & $1.032 \pm 0.046$ &  $8/7$ \\
$ \xi$ Planck$\times$1.042 & $1.018 \pm 0.041$ &  $7/9$\\
$\xi$ BPZ & $1.012 \pm 0.040$ &  $12/9$ \\
$\xi$ no $w_{\rm sys}$ & $1.029 \pm 0.040$ & $10/9$\\ 
$ \xi$ $\Sigma_{\rm nl}=4$ & $1.023 \pm 0.038$ &  $9/9$ \\
$ \xi$ $\Sigma_{\rm nl}=12$ & $1.043 \pm 0.052$ &  $11/9$ \\
$ \xi$ $\Sigma_{\rm nl}$ free & $1.024 \pm 0.039$ &  $9/9$ \\
$ \xi$ $A =0$ & $1.039 \pm 0.040$ &  $10/12$ \\
$\xi$ $0.7 < z < 1.0$ & $1.052 \pm 0.031$  & $17/9$ \\
\hline
\hline
\label{tab:baodata}
\end{tabular}
\end{table}

\clearpage

\subsection{BAO Measurements}

Here, we present the best-fit BAO results and likelihoods. Table \ref{tab:baodata} lists our BAO measurements for the DES Y1 data and robustness tests on these data. We find similar results for $w(\theta)$ and $\xi(s_{\perp})$, both in terms of $\alpha$ and its uncertainty. Fig. \ref{fig:BAOwxi} displays these results for our DES Y1 data using orange stars.
 for $0.6 < z < 1.0$ and yellow stars for $0.7 < z < 1.0$. Clearly, our results are consistent within the expected scatter.

Fig. \ref{fig:wthBAO} displays the BAO signal we measure in $w(\theta)$. To make this plot, we have subtracted the model obtained when using the same best-fit parameters but using the smooth $w(\theta)_{\rm noBAO}$ template (obtained from $P_{\rm nw}$). In order to plot each redshift bin clearly, we have added significant vertical offsets (and some small horizontal ones). One can see that the BAO feature in the model moves to lower values of $\theta$ as the redshift increases, as the co-moving location of the BAO feature is constant. Such a pattern is observed in the data for $z > 0.7$. The combination of these four $w(\theta)$ measurements, accounting for the covariance between the redshift bins, yields a measurement of $\alpha = 1.031\pm0.041$, i.e., approximately a 3 per cent greater angular diameter distance than predicted by our fiducial cosmology, but with 4 per cent uncertainty. The overall fit to the DES Y1 data is acceptable, as a $\chi^2=53$ for 43 degrees of freedom has a $p$-value of 0.14. Despite its appearance, the $0.6 < z < 0.7$ does not have a substantial affect on the goodness of fit; as the best-fit has a $\chi^2$/dof = 37/32 when these data are removed.

 Fig. \ref{fig:clbao} displays the Y1 BAO feature, isolated in
harmonic space and compared to the best-fit model. This figure is analogous to Fig.~\ref{fig:wthBAO} for
$w(\theta)$. Here, we see that the BAO feature in the model moves towards higher $\ell$ as the redshift increases and that this behavior is traced by the data points. In harmonic space we find $\alpha=1.023\pm0.047$, which is a shift in $\alpha$ of approximately 0.25$\sigma$ compared to the $w(\theta)$ measurement and slightly greater uncertainty for $C_{\ell}$. As shown in Fig \ref{fig:BAOwcl}, such differences are typical for the mock results. We are therefore satisfied with the agreement between configuration and harmonic space results. The $\chi^2/{\rm dof} = 94/63$ we obtain for the $C_{\ell}$ fit is slightly high. The formal $p$-value is 0.007, suggesting that the result is unlikely. Using an analytic covariance matrix (denoted `alt. cov' in Table \ref{tab:baodata}) instead of the one derived from mocks reduces the $\chi^2/{\rm dof} = 86/63$, with a $p$-value of 0.029 and shifts the result to $\alpha=1.039\pm0.053$. The combination of a significant bias on the results recovered from the mock analysis and the poor $\chi^2$ recovered from the fits to the DES Y1 data make us discount the use of the $C_{\ell}$ results as representing the signal in the DES Y1 data. However, the agreement with the $w(\theta)$ is encouraging as a robustness test and we expect future studies to make further use of the $C_{\ell}$ results given future methodological improvements. Further tests of the $C_{\ell}$ results can be found in \methodcl.

Fig. \ref{fig:clusBAO} displays the DES Y1 BAO signal in $\xi(s_{\perp})$ using $0.6 < z < 1.0$, again by subtracting the no BAO component of the best-fit model. This represents the result listed as `fiducial' in Table \ref{tab:baodata}. The $\chi^2=9$ for 9 dof. This result is chosen as the fiducial result $\xi$ (from among four choices of bin center) as it has very similar signal to noise and best fit value as the $w(\theta)$ result, which we will use for our DES Y1 measurement, and thus represents a highly compressed illustration of the DES Y1 BAO signal. The $\xi$ result we quote as `combined' in Table \ref{tab:baodata} is obtained from the mean likelihood of four $\xi$ results, each using a bin size of $12h^{-1}$Mpc with the bin size shifted in increments of 3$h^{-1}$Mpc; this procedure of taking the mean across the bin centers was demonstrated to slightly improve the results for mock data in Section \ref{sec:mocktest}. This result is similar to the $w(\theta)$ result, with a slightly greater uncertainty. Comparing the orange stars to the white circles in Fig \ref{fig:BAOwxi} indicates that the differences we find in the $w(\theta)$ and $\xi(s_{\perp})$ results are typical.

We recover, for both $w(\theta)$ and $\xi(s_{\perp})$, a smaller uncertainty when we ignore the $0.6 < z < 0.7$ data; i.e., the signal to noise appears greater in the $0.7 < z < 1.0$ sample than for the $0.6 < z < 1.0$ data. This is, of course, unexpected. In Appendix \ref{app:67}, we compare results obtained from mock realizations using both redshift ranges. We find that eight per cent of the realizations obtain an uncertainty that is improved by a greater factor than we find for DES Y1 when ignoring the $0.6 < z < 0.7$ data (and 30 per cent satisfy the condition that the $0.7 < z < 1.0$ uncertainty is less than the $0.6 < z < 1.0$ uncertainty). This eight per cent becomes more significant when one considers that to truly consider how likely the result is, we would have to test removing all independent equal sized volumes, not just those with $0.6 < z < 0.7$. These eight per cent of cases are thus not particularly unusual. Studying them further, we find that the $0.6 < z < 1.0$ results are more trust-worthy, but the uncertainty on $\alpha$ is likely over-estimated. Thus, we use the full $0.6 < z < 1.0$ data set for our DES Y1 result as this is the more conservative choice. A final decision to be made is how to treat the $w(\theta)$ and $\xi(s_{\perp})$ results. Given that the $w(\theta)$ results are more precise in the $0.6 < z < 1.0$ redshift range, are less biased when tested on the mock samples, and are less dependent on the choice of damping scale (see next subsection), we use the $w(\theta)$ results as our choice for the DES Y1 measurement.

\begin{figure}
\includegraphics[width=84mm]{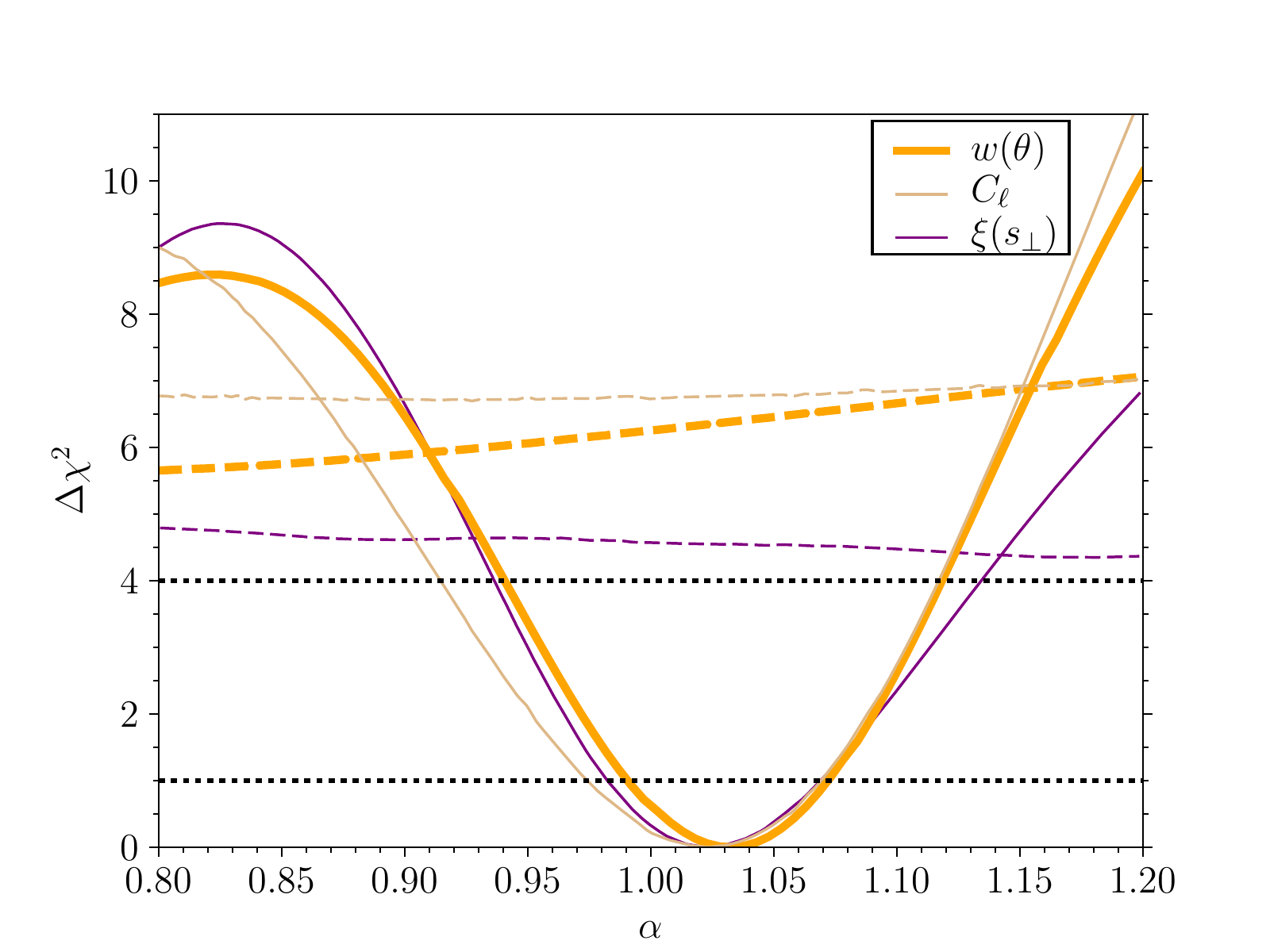}
  \caption{The BAO likelihood for DES Y1 data, for $w(\theta)$, $C_{\ell}$, and $\xi(s_{\perp})$. The dashed line shows the result for a model with no BAO, indicating that the data prefers a BAO feature at greater than 2$\sigma$ significance. The dotted black lines denote 1 and 2$\sigma$, based on $x\sigma=\sqrt{\Delta \chi^2}$.}
  \label{fig:BAOlik}
\end{figure}

Fig. \ref{fig:BAOlik} displays the $\Delta\chi^2$ likelihood for $\alpha$ using $w(\theta)$ and $\xi(s_{\perp})$. The dashed line is for the no BAO model (derived from $P_{\rm nw}$). We find a preference for BAO that is greater than 2$\sigma$ for both $w(\theta)$ and $\xi(s_{\perp})$. The $w(\theta)$ and $\xi(s_{\perp})$ likelihoods are close near the maximum likelihood, but diverge at high $\alpha$ values. Thus, our $\Delta \chi^2 = 1$ definition for $\alpha$ and its uncertainty recovers results that agree quite well. A summary of the differences is that $\xi$ rejects low $\alpha$ with slightly greater significance and $w(\theta)$ rejects high $\alpha$ with greater significance. We use the full $w(\theta)$ likelihood for any cosmological tests, as the Gaussian approximation is clearly poor outside of the $\sim1\sigma$ region.

\subsection{Robustness Tests}
\label{sec:robust}

We vary our methodology in a variety of ways in order to test the robustness of our results. We have already shown that $\xi$ and $w(\theta)$ obtain consistent results and that the change in results when eliminating $0.6 < z < 0.7$ data are consistent with expectations based on tests on mock realizations. Here, we consider how the results change with different modeling assumptions, changes in the range of scales used, the bin size, the use of the systematic weights, the choice of photometric redshifts, and if a harmonic-space estimator is used instead of the configuration space ones. The results are tabulated in the bottom rows of Table \ref{tab:baodata}.

One significant concern in our work is the use of photometric redshifts. Throughout, we have assumed the fiducial $\phi(z_{\rm true})$ results obtained from the redshift validation described in \sample ~are correct. However, the impact of this validation on our recovered results is relatively small. The output of DNF provides an initial estimate of the redshift distribution, which we also used to produce $w(\theta)$ BAO templates. Using these BAO templates, labeled `z uncal' in Table \ref{tab:baodata}, our measurement by only $0.25\sigma$ towards a smaller $\alpha$ value. This test performed only for $w(\theta)$, but given that the main effect is simply to shift the mean redshifts assumed in the analysis, we expect similar results for the other statistics. Further, this test presents the results obtained with no attempt to calibrate the photometric redshifts and instead simply stacking the output from the redshift pipeline. It shows that our results shift by only a fraction of a $\sigma$. The redshift validation produces more accurate determinations of the redshift distributions. The estimated uncertainties in the mean redshift obtained from the process are less than 0.8 per cent at all redshifts, i.e., $\sigma(\bar{z})/\bar{z} < 0.008$. The effect on the BAO measurement can be approximated as this being a constant shift in all redshifts and this affects each clustering estimator equally. This converts to a systematic uncertainty on our $\alpha$ measurement of 0.006. This is 15 per cent of our statistical uncertainty and thus negligible for this DES Y1 analysis. More detailed tests are shown in \methodACF. 

An additional test of the robustness to redshift uncertainties is to use a different method, BPZ, to estimate the photometric redshifts and reproduce the sample and measurements. For this test, we still use the covariance obtained from the fiducial 1800 
mock realizations with redshift distributions matched to our fiducial DNF photometric redshifts. However, the BAO templates for the measurements are produced using the BPZ redshift distributions estimated through their redshift validation.
As for DNF, we find similar results for $w(\theta)$ and $\xi$, as each shifts to a slightly smaller value of $\alpha$. This is a 0.32$\sigma$ shift for $\xi$ and 0.37$\sigma$ for $w(\theta)$.
For $w(\theta)$, this is the level of shift expected if the two samples have a correlation, $c$, of 0.93 (calculated via $\sigma = \sqrt{2-2*c}$).
Given that only 88 per cent of the galaxies used are the same (it is less than 100 per cent because galaxies scatter across the redshift boundaries) and the number of matches per $\Delta z = 0.1$ redshift bin is at most 56 percent, we are confident the expected correlation in BAO results obtained from the two redshift estimates is less than 0.93. We therefore conclude the differences in the results between the two redshift estimates are not statistically significant.


The damping scale that we assume in our BAO template affects our results, mostly in terms of the recovered uncertainty for $w(\theta)$ (which we use for our Y1 measurement). For both $w(\theta)$ and $\xi$, we test both halving our fiducial scale and making it 50 per cent greater. This change is far greater than the change expected between our fiducial cosmology and that of \cite{Planck2015}. For $w(\theta)$, using the low damping scale shifts $\alpha$ lower by 0.005 (0.12$\sigma$) and decreases the uncertainty by 15 per cent. Using the greater damping scale does not shift $\alpha$ but increases the uncertainty by 37 per cent. The data slightly prefer the lower damping scale, as $\Delta\chi^2=2$ compared to the fiducial case. Allowing the damping to be free has a best-fit at low $\Sigma_{\rm nl}$ values and thus significantly reduces the uncertainty but shifts $\alpha$ (lower) by only 0.12$\sigma$. Thus, the distance obtained from our $w(\theta)$ results used for our DES Y1 measurement is robust to the choice of damping scale. The uncertainty on the measurement depends fairly strongly on how the damping scale is treated, and we thus choose to use the damping scale that is best-fit to our mock samples, $\Sigma_{\rm nl} = 5.2 h^{-1}$Mpc, which is close to theoretical expectations. 

For $\xi(s_{\perp})$, we find a similar effect on the uncertainty and greater shifts in the recovered $\alpha$. The $\alpha$ values shift by $\sim \pm 0.25\sigma$. Allowing $\Sigma_{\rm nl}$ to be a free parameter yields a result that is close to the result for $\Sigma_{\rm nl} = 4 h^{-1}$Mpc, as, similar to $w(\theta)$, the DES Y1 data slightly prefers the smaller damping scale. Thus, marginalizing over $\Sigma_{\rm nl}$ results in a smaller estimated uncertainty. Therefore, as with $w(\theta)$, we choose to use the results with the fixed damping scale as our measurement. However, for the DES Y1 data set the $\xi(s_{\perp})$ results are not as robust to the choice of damping value as $w(\theta)$, which is one of the reasons we use the $w(\theta)$ results as our choice for the DES Y1 measurement.

When ignoring the systematic weights, we find almost no change in the recovered BAO measurements. The change is greatest for $w(\theta)$, but only 0.13$\sigma$. This is consistent with results from spectroscopic surveys \citep{Ross16,Ata17}, which have consistently demonstrated that BAO measurements are robust to observational systematic effects. Additionally, no large deviations are found when the bin size or range of scale fit are changed. Finally, there is no significant change in our results if we set the broadband polynomial terms to 0, denoted by `$A=0$' in the Table \ref{tab:baodata}.

As a final test, we obtain results when we assume a cosmology consistent with Planck $\Lambda$CDM \citep{Planck2015} for calculating both the pair counts and BAO template. For this, we use the same fiducial cosmology recently assumed in \cite{Acacia}, which is flat $\Lambda$CDM with $\Omega_{\rm matter} = 0.31$, $h=0.676$, and $\Omega_{\rm baryon}h^2=0.022$. When using this cosmology, we expect to obtain an $\alpha$ value that is lower by 1.042, thus in Table \ref{tab:baodata} we have multiplied the Planck results by 1.042. For $\xi$, the shift is 0.008 in $\alpha$ ($0.18\sigma$) away from the combined result using the MICE cosmology. For $\xi$, we expect a difference similar to what was found for the bin center tests, as the pair-counts were re-calculated assuming the Planck cosmology; our result is only slightly greater than the variance associated with alternative binnings. For $w(\theta)$, the results using the Planck template are nearly identical to those using the default MICE template as all differences are within 0.002. For $w(\theta)$, we also alter the covariance matrix so that the Gaussian part is given by the \cite{Planck2015} cosmology, using Eq. \ref{eq:newcov}. Thus, we conclude the choice of fiducial cosmology has a negligible effect on our results.

\section{Cosmological Context}
\label{sec:cosmo}
Our DES Y1 BAO measurement can be used to constrain cosmological models, given the likelihood for $\alpha$ (shown in Fig. \ref{fig:BAOlik}), our fiducial cosmology, and the effective redshift of our measurement. This requires multiplying the $\alpha$ measurements by $\frac{D^{\rm fid}_A(z_{\rm eff})}{r^{\rm fid}_{\rm d}} = 10.41$ and testing this against the $\frac{D_A(z_{0.81})}{r_{\rm d}}$ predicted by any given cosmological model\footnote{We will make our likelihood publicly available after this work has been accepted for publication by the journal.}. In the top row of Table \ref{tab:baodata}, we have thus multiplied the $w(\theta)$ results by 10.41 in order to quote our Y1 measurement of $\frac{D_A(z_{0.81})}{r_{\rm d}} = 10.75\pm0.43$.

\begin{figure}
\includegraphics[width=84mm]{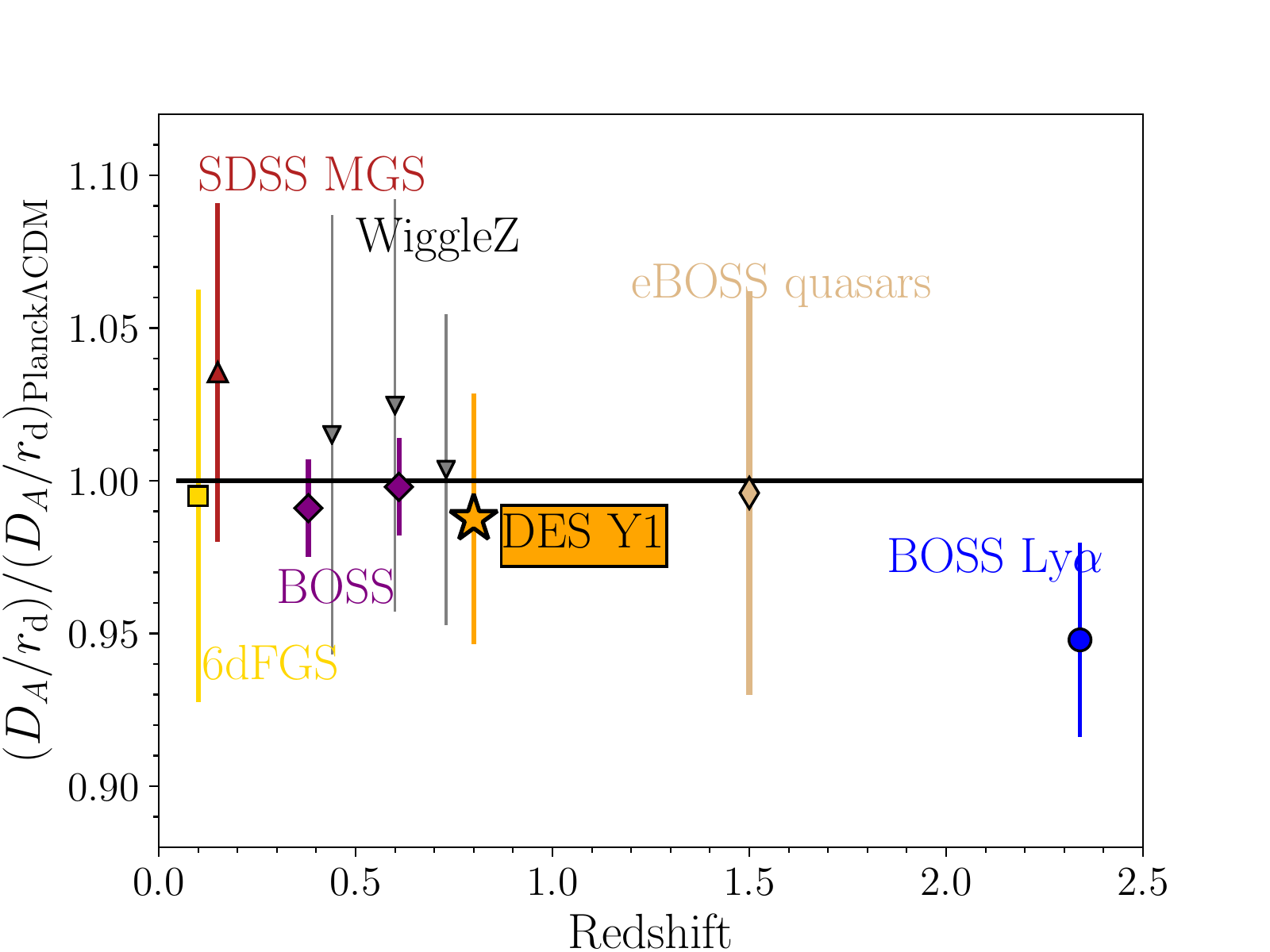}
  \caption{Measurement of the angular diameter distance measured from BAO, compared to the Planck $\Lambda$CDM prediction. The DES Y1 measurement is shown using a gold star. The additional measurements are described in the text. Where possible, we have used the available $D_A$ measurement. Some studies have reported only spherically averaged distances; in these cases we have multiplied the uncertainty by 1.5 (as this is the approximate scaling between spherically averaged and transverse signal to noise). The results where this scaling was applied are 6dFGS, SDSS MGS, and eBOSS quasars.
  }
  \label{fig:BAOladder}
\end{figure}

Fig. \ref{fig:BAOladder} displays our measurement, compared to other BAO angular diameter distance measurements and the Planck $\Lambda$CDM prediction (with fixed minimal neutrino mass). We include measurements from \cite{6dF} (6dFGS), \cite{Ross15} (SDSS MGS), \cite{Acacia} (BOSS), \cite{Kazin14} (WiggleZ), \cite{Ata17} (eBOSS quasars) and the combination of \cite{Bautista17} and \cite{DR12lyax} (BOSS Ly$\alpha$). These make up the most up to date, and largely independent, BAO distance ladder. Many of the BAO measurements were made in terms of the spherically averaged distance quantity, which is a combination of the angular diameter distance and $H(z)$. Assuming spherical symmetry, the $D_A(z)$ constraints are 50 per cent less precise \citep{Ross152D} and we have thus multiplied the error-bars by 1.5 while fixing the relative $(D_A/r_{\rm d})/(D_A/r_{\rm d})_{\rm Planck\Lambda CDM}$ value to be the same as for spherically averaged measurement.

The DES Y1 measurement is consistent with expectations from Planck $\Lambda$CDM and previous BAO measurements. While the new measurement is not nearly as precise as BOSS measurements, it is the most precise $D_A$ measurement in the range $0.6 < z < 2$. The DES Y1 and the 6dFGS (yellow square) are the only measurements displayed that do not rely on imaging from SDSS. The DES measurement is the only one that does not rely on spectroscopic redshifts. The precision of our measurement is similar to that obtained previously at $z\sim 0.55$ by the combination of \cite{Crocce11} and \cite{Carnero12} and independently \cite{Seo12} using SDSS imaging data and photometric redshifts (not plotted).

We can also compare with the recent DES Y1 results obtained in \citealt{Abbot173x2pt} (hereafter `DES 3x2pt'). Converting the results from the DES 3x2pt $\Lambda$CDM Monte-Carlo Markov Chains to a posterior likelihood for $D_A(0.81)/r_{\rm d}$, we obtain 10.59$\pm$0.44. Thus, our DES Y1 BAO measurement is in agreement with the results of DES 3x2pt assuming the $\Lambda$CDM model.

\section{Conclusions}

The results of this study can be summarized as follows:

$\bullet$ We have used a sample of 1.3 million DES Y1 galaxies spread over 1336deg$^2$ with $0.6 < z_{\rm photo} < 1.0$, defined in \cite{Crocce17} (\sample), in order to obtain a four per cent measurement of the ratio of the angular diameter distance to $z=0.81$ to the size of the BAO standard ruler set by early Universe recombination physics: $D_A(z_{\rm eff}=0.81)/r_{\rm d} = 10.75\pm 0.43 $.

$\bullet$ In order to construct covariance matrices and set analysis choices, we have used 1800 mock realizations of the DES Y1 sample, constructed as described in \cite{Avila17} (\mocks). Our DES Y1 results are typical given the distribution of recovered from the 1800 mock realizations.

$\bullet$ We have used three separate projected clustering statistics, one defined in terms of the angular separation ($w(\theta)$, with methodology described in \citealt{whetabao}; \methodACF), another defined in terms of the projected physical separation ($\xi(s_{\perp})$, with methodology described in \citealt{RossOpt3D}; \methodXi), and a third in spherical harmonic space ($C_{\ell}$, with methodology described in \citealt{clbao}; methodcl). Each statistic returns consistent results (the difference is 0.25$\sigma$ in the best-fit value and the uncertainties differ by less than 15 per cent). These differences are consistent with differences found in the mock realizations. The preference for a BAO feature in the data is greater than 2$\sigma$ for both clustering statistics. We use the $w(\theta)$ result as our consensus DES Y1 measurement, as its methodology was demonstrated to be most accurate when testing against mocks and it was more robust when testing the results obtained from various treatment of the DES Y1 data.

$\bullet$ We obtain results using two photometric redshift estimates, one machine learning based (DNF) and the other template based (BPZ), and obtain results that are matched to within $0.37\sigma$. The DES Y1 results are based on the DNF results, which are the most precise and accurate as described, along with their validation in \sample. The results from the validation suggest that uncertainties in the accuracy of the DNF redshifts contribute a negligibly small systematic uncertainty to our DES Y1 BAO measurements.

$\bullet$ We find no significant changes in the $w(\theta)$ BAO measurements when varying the methodology, including omitting corrections for observational systematics and using a different fiducial cosmology.

In short, we find our DES Y1 BAO measurement is robust to a number of stress tests and is consistent with our simulations of the DES Y1 dataset. These results are the first BAO measurement obtained from DES. DES year three (Y3) data has already been observed and occupies the full 5000 deg$^2$ with greater average coverage in the number of exposures than Y1 over our 1336 deg$^2$. With these data, we expect to obtain results with approximately a factor of two smaller statistical uncertainty in the near future. 

 As a byproduct of the preparation for this analysis we have identified a number of items where improvement or further study will benefit DES Y3 measurements: \begin{itemize}
\item Include a more realistic modeling of photometric redshift errors in the template of $\xi(s_\perp)$, which currently assumes gaussian errors, which we believe are the reason for the systematic bias of $0.13\sigma$ when using that statistics, as discussed in Sec. \ref{sec:testonmean} and shown in Table \ref{tab:baomean}  
\item Reduce the uncertainty in the mean of the redshift distributions, which currently represents a systematic uncertainty of $0.15 \sigma$, see Sec. \ref{sec:robust}; this is the greatest of any systematic uncertainty we have identified that affects $w(\theta)$ results  
\item Revisit the theoretical template for the $C_\ell$ methodology, or otherwise pin down the cause of the biases shown in Table \ref{tab:baomean}
\item Further investigate the dependence of the covariance matrix on cosmological parameters and other choices (e.g., the type of mock used). 
\end{itemize}
Altogether, based on our results, none of the above items will lead to systematic uncertainties close to the future level of statistical precision.

We expect our work can be used as a guide for future imaging surveys such as the Large Synoptic Survey Telescope (LSST; \citealt{lsst}).
In turn, BAO measurements had previously been proven to be a robust and precise method for measuring cosmological distances when using spectroscopic redshifts (\citealt{Acacia} and references there-in). We have obtained a similar level of robustness using a purely photometric data set. 
Based on this work, we expect DES Y3 data to provide the most precise BAO angular diameter distance measurement, excluding BOSS galaxy results. Further, the increased photometric depth of the Y3 and then final DES data will allow extensions to higher redshifts than probed here. Thus, together with emission line galaxy data from the extended Baryon Oscillation Spectroscopic Survey (eBOSS; \citealt{eboss}; comparable precision is expected), these data will be the first to use galaxies beyond redshift 1 to measure BAO. These measurements will thus pave the way for those obtained by the Dark Energy Spectroscopic Instrument (DESI;\citealt{DESI1,DESI2}) and Euclid \citep{Euclid11,Euclid}.

\section*{Acknowledgments}

AJR is grateful for support from the Ohio State University Center for
Cosmology and AstroParticle Physics. MC acknowledges support from the
Spanish Ramon y Cajal MICINN program.
KCC acknowledges the support from the Spanish Ministerio de Economia y Competitividad grant ESP2013-48274-C3-1-P and the Juan de la Cierva fellowship. NB acknowledges the use of University of Florida's supercomputer HiPerGator 2.0 as well as thanks the University of Florida's Research Computing staff. HC is supported by CNPq. ML and RR are partially supported by FAPESP and CNPq. We are thankful for the support of the Instituto Nacional de Ci\^encia
e Tecnologia (INCT) e-Universe (CNPq grant 465376/2014-2).

This work has made use of CosmoHub, see \cite{Carretero17}. CosmoHub has been developed by the Port d'Informaci\'{o} Cient\'{i}fica (PIC), maintained through a collaboration of the Institut de F\'{i}sica d'Altes Energies (IFAE) and the Centro de Investigaciones Energ\'{e}ticas, Medioambientales y Tecnol\'{o}gicas (CIEMAT), and was partially funded by the ``Plan Estatal de Investigaci\'{o}n Científica y T\'{e}cnica y de Innovaci\'{o}n'' program of the Spanish government.

We are grateful for the extraordinary contributions of our CTIO colleagues and the DECam Construction, Commissioning and Science Verification
teams in achieving the excellent instrument and telescope conditions that have made this work possible.  The success of this project also 
relies critically on the expertise and dedication of the DES Data Management group.

Funding for the DES Projects has been provided by the U.S. Department of Energy, the U.S. National Science Foundation, the Ministry of Science and Education of Spain, 
the Science and Technology Facilities Council of the United Kingdom, the Higher Education Funding Council for England, the National Center for Supercomputing 
Applications at the University of Illinois at Urbana-Champaign, the Kavli Institute of Cosmological Physics at the University of Chicago, 
the Center for Cosmology and Astro-Particle Physics at the Ohio State University,
the Mitchell Institute for Fundamental Physics and Astronomy at Texas A\&M University, Financiadora de Estudos e Projetos, 
Funda{\c c}{\~a}o Carlos Chagas Filho de Amparo {\`a} Pesquisa do Estado do Rio de Janeiro, Conselho Nacional de Desenvolvimento Cient{\'i}fico e Tecnol{\'o}gico and 
the Minist{\'e}rio da Ci{\^e}ncia, Tecnologia e Inova{\c c}{\~a}o, the Deutsche Forschungsgemeinschaft and the Collaborating Institutions in the Dark Energy Survey. 

The Collaborating Institutions are Argonne National Laboratory, the University of California at Santa Cruz, the University of Cambridge, Centro de Investigaciones Energ{\'e}ticas, 
Medioambientales y Tecnol{\'o}gicas-Madrid, the University of Chicago, University College London, the DES-Brazil Consortium, the University of Edinburgh, 
the Eidgen{\"o}ssische Technische Hochschule (ETH) Z{\"u}rich, 
Fermi National Accelerator Laboratory, the University of Illinois at Urbana-Champaign, the Institut de Ci{\`e}ncies de l'Espai (IEEC/CSIC), 
the Institut de F{\'i}sica d'Altes Energies, Lawrence Berkeley National Laboratory, the Ludwig-Maximilians Universit{\"a}t M{\"u}nchen and the associated Excellence Cluster Universe, 
the University of Michigan, the National Optical Astronomy Observatory, the University of Nottingham, The Ohio State University, the University of Pennsylvania, the University of Portsmouth, 
SLAC National Accelerator Laboratory, Stanford University, the University of Sussex, Texas A\&M University, and the OzDES Membership Consortium.

Based in part on observations at Cerro Tololo Inter-American Observatory, National Optical Astronomy Observatory, which is operated by the Association of 
Universities for Research in Astronomy (AURA) under a cooperative agreement with the National Science Foundation.

The DES data management system is supported by the National Science Foundation under Grant Numbers AST-1138766 and AST-1536171.
The DES participants from Spanish institutions are partially supported by MINECO under grants AYA2015-71825, ESP2015-66861, FPA2015-68048, SEV-2016-0588, SEV-2016-0597, and MDM-2015-0509, 
some of which include ERDF funds from the European Union. IFAE is partially funded by the CERCA program of the Generalitat de Catalunya.
Research leading to these results has received funding from the European Research
Council under the European Union's Seventh Framework Program (FP7/2007-2013) including ERC grant agreements 240672, 291329, and 306478.
We  acknowledge support from the Australian Research Council Centre of Excellence for All-sky Astrophysics (CAASTRO), through project number CE110001020.

This manuscript has been authored by Fermi Research Alliance, LLC under Contract No. DE-AC02-07CH11359 with the U.S. Department of Energy, Office of Science, Office of High Energy Physics. The United States Government retains and the publisher, by accepting the article for publication, acknowledges that the United States Government retains a non-exclusive, paid-up, irrevocable, world-wide license to publish or reproduce the published form of this manuscript, or allow others to do so, for United States Government purposes.

\appendix

\section{Comparison between $0.6 < z < 0.7$ and $0.7 < z < 1.0$ results}
\label{app:67}

\begin{figure}
\includegraphics[width=84mm]{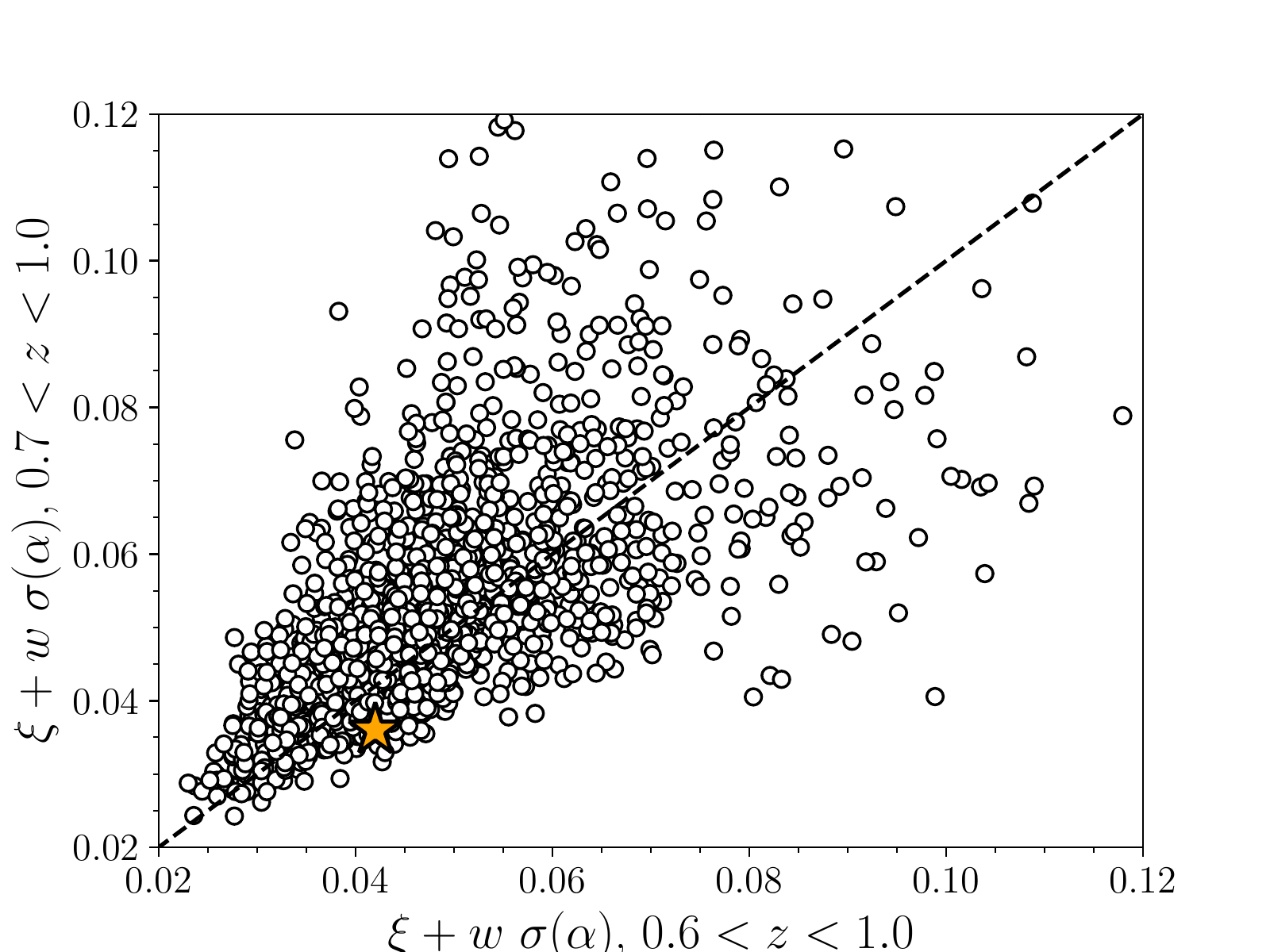}
\includegraphics[width=84mm]{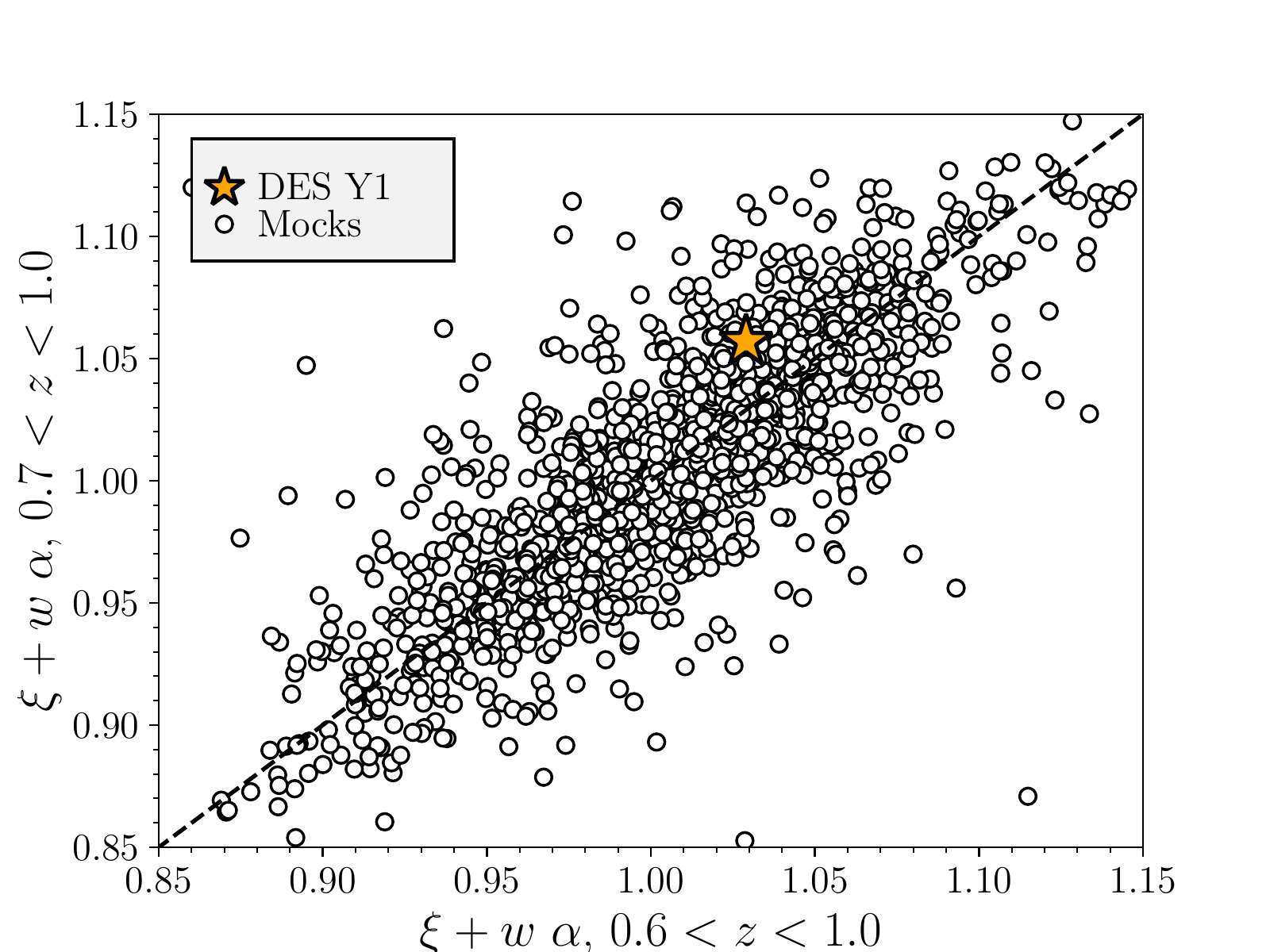}
  \caption{A comparison of $\xi+w$ BAO fit results when using the full $0.6 < z < 1.0$ redshift range and one restricted to $0.7 < z < 1.0$. 
  }
  \label{fig:BAO67}
\end{figure}

As described in Section \ref{sec:results}, we obtain more precise BAO measurements when excluding data with $0.6 < z < 0.7$. In this appendix, we use the mock realizations in order to assess how unusual this is and how best to treat the results in such a case. We compare the results of fitting to 1800 mock realizations without using the $0.6 < z < 0.7$ data compared to the full range. This is shown in Fig. \ref{fig:BAO67}. 1538 mock realizations (85 per cent) have a detection in both redshift ranges. As to be expected the results are strongly correlated. We see that the recovered uncertainty is usually greater when omitting the $0.6 < z < 0.7$, but this is not always the case. We find 130 (8 per cent) of the mock realizations recover an uncertainty $\sigma_1/\sigma_2 > 0.043/0.036$. The values $\sigma=0.43$ and 0.36 represent the mean of the $w(\theta)$ and $\xi(s_{\perp})$ uncertainties recovered from the DES Y1 data for the respective redshift ranges. Thus, our data measurement is not particularly unlikely in terms of this statistic, even without folding in the look-elsewhere effect, which would further decrease the significance because we are only considering a single anomalous statistic.

As can be seen  based on the position of the orange star in the bottom panel of Fig. \ref{fig:BAO67}, the value of $\alpha$ shifts when removing the $0.6 < z < 0.7$ data, but remains within the locus of mock realization results. In order to quantify the shift, we consider $|\alpha_1-\alpha_2|/\sigma_1$ (using 1 to denote $0.6 < z < 1.0$ and 2 to denote $0.7 < z < 1.0$) for each mock realization. We find the quantity is greater than that of the DES Y1 data in 443 (29 per cent) of the mock realizations. Thus, the difference in $\alpha$ is not unusual. 

At the end of Section \ref{sec:mocktest}, we found a correlation between the mean uncertainty and the scatter in $\alpha$ that suggested the obtained BAO likelihoods can generally be trusted. If this were to be true in all cases, it would suggest we should use only the $0.7 < z < 1.0$ data for our DES Y1 measurement. To test this, we further consider the 8 per cent of mock realizations that exhibit more extreme behavior in the ratio of uncertainties with and without removing the $0.6 < z < 0.7$ data. These realizations have been selected in order to have significantly greater uncertainty for the $0.6 < z < 1.0$ realizations and we indeed obtain $\langle \sigma_1 \rangle = 0.073$ compared to $\langle \sigma_2 \rangle = 0.054$. However, the standard deviations are flipped, $S_1 = 0.054$ compared to $S_2 = 0.061$. This suggests that in these cases the uncertainty is significantly over-(under-) estimated for the $0.6 < z < 1.0$ ($0.7 < z < 1.0$) and that the under-estimated $0.7 < z < 1.0$ uncertainty is actually a better match to the uncertainty of the $0.6 < z < 1.0$ results. In order to consider cases that are even more similar to our DES Y1 case, we take only the mocks out of this eight per cent that have an uncertainty less than 0.05 for $0.6 < z < 1$. This yields only 22 mocks. Their mean uncertainties are $\langle \sigma_1 \rangle = 0.046$ and $\langle \sigma_2 \rangle = 0.037$, to be compared to standard deviations of $S_1 = 0.047$ and $S_2 = 0.046$. While this is a limited number of mocks, the results are consistent with the conclusion that the uncertainty is under-estimated in the cases where $0.7 < z < 1.0$ yields less uncertainty than the $0.6 < z < 1.0$ data.

The findings can be summarized as
\begin{itemize}
\item Eight per cent of the mock realizations are more extreme in terms of the comparison between the recovered $0.6 < z < 1.0$ and $0.7 < z < 1.0$ uncertainties.
\item These realizations have been chosen to have worse uncertainties for $0.6 < z < 1.0$; we find $\langle \sigma_{\alpha} \rangle = 0.073$ for $0.6 < z < 1.0$ and $\langle \sigma_{\alpha} \rangle = 0.054$ for $0.7 < z < 1.0$.
\item For these eight per cent of realizations, we find the standard deviations for the scatter in the recovered $\alpha$ values are 0.054 for $0.6 < z < 1.0$ and 0.061 for $0.7 < z < 1.0$.
\item Changing our criteria to be only cases where the $0.6 < z < 1.0$ uncertainty is less than 0.05 (and thus similar to our DES Y1 result) or simply $\sigma_2/\sigma_1 > 1$ (which is true for 30 per cent of the mock cases) yields consistent results.
\end{itemize}

Thus, for these cases, the uncertainties are generally under-estimated for $0.7 < z < 1.0$ and we consider the $0.6 < z < 1.0$ result more trustworthy.

\label{lastpage}

\end{document}